\documentclass[journal]{IEEEtran}

\usepackage[whole]{bxcjkjatype}
\usepackage[utf8]{inputenc}
\usepackage{cite}
\usepackage{amsmath,amssymb,amsfonts}
\usepackage{algorithmic}
\usepackage{graphicx}
\usepackage{textcomp}
\usepackage[varg]{newtxmath}
\usepackage{algorithm}
\usepackage{setspace}
\usepackage{epsfig}
\usepackage{multirow}
\usepackage{bigstrut}
\usepackage{color}
\usepackage{subcaption}

\def\x{{\mathbf x}}

\newcommand{\argmin}{\mathop\textrm{argmin}\limits}

\newcommand{\bvec}[1]{\mbox{\boldmath $#1$}}
\newcommand{\unfold}{\textrm{unfold}}

\newcommand{\proj}{\textrm{proj}}

\def\k{{\mathbf k}}
\def\l{{\mathbf l}}

\def\r{{\mathbf r}}

\def\w{{\mathbf w}}
\def\x{{\mathbf x}}
\def\y{{\mathbf y}}

\def\X{{\mathbf X}}
\def\Y{{\mathbf Y}}
\def\cS{{\mathcal S}}
\def\cV{{\mathcal V}}
\def\cX{{\mathcal X}}
\def\cY{{\mathcal Y}}
\def\cZ{{\mathcal Z}}
\def\A{{\mathbf A}}

\def\V{{\mathbf V}}
\def\Z{{\mathbf Z}}
\def\U{{\mathbf U}}

\def\R{{\mathbb R}}

\def\BibTeX{{\rm B\kern-.05em{\sc i\kern-.025em b}\kern-.08em
    T\kern-.1667em\lower.7ex\hbox{E}\kern-.125emX}}

\begin{document}
\title{On The Synergy Between Nonconvex Extensions of The Tensor Nuclear Norm for Tensor Recovery}
\author{\uppercase{Kaito Hosono}, \IEEEmembership{Member, IEEE},
\uppercase{Shunsuke Ono}, \IEEEmembership{Member, IEEE}, 
and \uppercase{Takamichi Miyata}., \IEEEmembership{Member, IEEE}
\thanks{K. Hosono and T. Miyata are with the Chiba Institute of Technology, Narashino-shi 275-0016 Japan.}
\thanks{S. Ono was with the Department of Computer Science, School of Computing, Tokyo Institute of Technology, Meguro-ku 152-8552 Japan.}
\thanks{This work was supported by JSPS KAKENHI Grant Number JP19K04377.}
}

\maketitle

\begin{abstract}
Low-rank tensor recovery has attracted much attention among various tensor recovery approaches.
A tensor rank has several definitions, unlike the matrix rank--e.g. the CP rank and the Tucker rank.
Many low-rank tensor recovery methods are focused on the Tucker rank.
Since the Tucker rank is nonconvex and discontinuous, many relaxations of the Tucker rank have been proposed, e.g., the tensor nuclear norm, weighted tensor nuclear norm, and weighted tensor Schatten-$p$ norm.
In particular, the weighted tensor Schatten-p norm has two parameters, the weight and $p$, and the tensor nuclear norm and weighted tensor nuclear norm are special cases of these parameters.
However, there has been no detailed discussion of whether the effects of the weighting and $p$ are synergistic.
In this paper, we propose a novel low-rank tensor completion model using the weighted tensor Schatten-$p$ norm to reveal the relationships between the weight and $p$.
To clarify whether complex methods such as the weighted tensor Schatten-$p$ norm are necessary, we compare them with a simple method using rank-constrained minimization.
It was found that the simple methods did not outperform the complex methods unless the rank of the original tensor could be accurately known.
If we can obtain the ideal weight, $p = 1$ is sufficient, although it is necessary to set $p<1$ when using the weights obtained from observations.
These results are consistent with existing reports.
\end{abstract}


% \fi
\IEEEpeerreviewmaketitle

\section{Introduction}
\label{sec:intro}
%テンソルとは
%テンソルとは，多次元的な情報やそれらの関係性をよく表現できる強力なツールであり，画像処理，信号処理分野で広く用いられている．
A tensor is a powerful tool that can describe multidimensional information and the complex relationships among elements, and it is widely used in the field of signal and image processing\cite{Gandy2011,Roughan2012,Liu2013,Liu2016,Ng2017,Sun2018,Xie2018,Kong2018,Zhang2019,Yokota2019,Wang2019,Hosono2019}.
%通常はこれらの情報を観測から完璧に得ることはできない
Usually, such information cannot be fully obtained through observation, and we need to complete or recover a full tensor from incomplete or degraded measurements, which are corrupted by noise, missing entries, and/or outliers.
Among various tensor completion/recovery approaches, low-rank-based methods have attracted much attention because they exploit the essential structure of tensors and achieve accurate estimation.
%テンソル復元とは
%テンソル復元は，ノイズや欠損などにより劣化したテンソルから，元のテンソルを推定するタスクである．
%A goal of tensor restoration is to estimate a clean original tensor from a degraded tensor, which is corrupted by added noise, missing entries, and other degradation factors. 
%このとき，元のテンソルに関する何らかの事前情報を利用することで，高精度な復元が可能となる．
%事前情報として特に注目されているのが，テンソルのランクの小ささ（低ランク性）である．
%Recently, low-rankness prior has attracted much attention to recover accurate tensors from the corresponding corrupted observations. 
%テンソル復元のうち，元のテンソルに低ランク性を仮定するものを低ランクテンソル復元と呼ぶ．
%A category of tensor restoration method, which assumes the  low-rankness of original tensors, is called low-rank tensor restoration.

%LRTCは大事
%行列とは異なり，テンソルのランクには，複数の異なった定義が存在する．
%なかでも，CPランクとTuckerランクの2種類がよく知られている．
Unlike the matrix rank, there are several different definitions of the tensor rank;  well-known examples include the CANDECOMP/PARAFAC (CP) rank\cite{Harshman1970} and the Tucker rank\cite{Tucker1966}.
%近年では，Tuckerランクに着目した低ランクテンソル復元が数多く提案されている[横田ら，HaLRTCとかもここ？]．
Since determining the CP rank is NP-hard\cite{Hastad1990}, many existing low-rank tensor recovery methods are focused on the Tucker rank.
%Tuckerランクはテンソルを行列へと変換する展開操作と，行列ランクを組み合わせたものであり，n階テンソルのTuckerランクはn次元ベクトルとなる．
The Tucker rank is defined as follows: 1) an input tensor is converted to the matrices (by the unfolding operation); 2) the average rank of these matrices is calculated.
%平均Tuckerランクは行列ランクを用いているため，非凸かつ不連続な関数であるという問題がある．
The Tucker rank is very difficult to handle because of its nonconvexity and discontinuity.

%TNNファミリー
%この問題を解決するため，平均Tuckerランクの凸緩和であるテンソル核ノルムを用いる手法が提案された[Gandyら，Liuら]．
To address this problem, the tensor nuclear norm, which is a convex surrogate of the Tucker rank, is proposed\cite{Gandy2011,Liu2013}.
%テンソル核ノルムで用いられる核ノルムは特異値の絶対値和として定義されており，行列ランクの凸緩和であることが知られている連続関数である．
Methods based on the Tucker rank replace the rank of unfolding matrices with their nuclear norms, where the nuclear norm is known as a continuous tightest convex surrogate of the matrix rank\cite{Fazel2002}.

%一方で，行列の低ランク性の指標として，核ノルムを一般化した重み付き核ノルム[Chenら]や，schatten-pノルム[]が提案されている．
On the other hand, the weighted nuclear norm and the Schatten-$p$ norm have been proposed as different surrogates of the matrix rank\cite{Chen2013,Xie2016,Gu2017,Zha2018}.
%これらの手法は，低ランクな行列の復元問題において核ノルムと比べて高い性能を発揮することが知られている．
Both are a generalization of the nuclear norm and usually perform better than the nuclear norm for low-rank matrix recovery. 
%重み付き核ノルム，schatten-pノルムをテンソルへと拡張した重み付きテンソル核ノルム，テンソルschatten-pノルムが提案されている．
%重み，pの決め方が大事という部分
%どちらの手法もテンソル核ノルムと比べて高精度な低ランクテンソル復元が可能だが，重みやパラメータpを適切に決定する必要がある．
Following this trend, a weighted tensor nuclear norm and a tensor Schatten-$p$ norm have also been proposed\cite{Hosono2019,Kong2018}.
They are extensions of the weighted nuclear norm and the Schatten-$p$ norm for tensors, respectively, and they generally perform better for low-rank tensor recovery as well.
However, for effective use, we need to select appropriate weights and parameters $p$.

%重み，pの決定方法に関する話
%重み付き核ノルムでよく用いられる重みとして，原行列特異値の逆数（オラクル重み）がある．
The ideal (oracle) weights for the weighted nuclear norm are the inverses of the singular values of the original matrix.
%オラクル重みが用いられるのは，原行列の重み付き核ノルムが原行列のランクと一致するためである．
This is because the weighted nuclear norm with the oracle weights of the original matrix is identical to the rank.
%一般に原行列の特異値を知るのは困難であるため，多くの場合原行列を何らかの方法で推定し，これを用いて重みを決定する．
Generally, obtaining the singular values of the original matrix is difficult. 
Therefore, for practical usage, we need some methods to estimate the singular values of the original matrix to determine the 
%weight.
%Since obtaining the singular values of the original matrix is difficult, we need to estimate the singular values of the original matrix to someway to determine the 
weights\cite{Gu2017}.
%schatten-pノルムのパラメータpは，発見的に決定されることが多く，殆どの場合$p<1$が用いられる．
On the other hand, the parameter $p$ for the Schatten-$p$ norm is generally determined in a heuristic manner and in most cases, $p<1$ is employed\cite{Xie2016,Kong2018,Zhang2019}.
We should note that both the weighted tensor nuclear norm and the tensor Schatten-$p$ norm (with $p<1$) are in general nonconvex, as is the case with the matrix counterparts.
%画像復元での応用では，$p\simeq 0.9$が最も性能が高いとする報告がされている．
%Some previous studies revealed that $p\simeq 0.9$ performs well for image restoration.

%平均tuckerランクは行列ランクを用いているため，非凸かつ不連続な関数であるという問題がある．
%この問題を解決するため，平均Tuckerランクの凸緩和であるテンソル核ノルムを用いる手法が提案された[Gandyら，Liuら]．
%核ノルムは特異値の絶対値和として定義されており，行列ランクの凸緩和であることが知られている連続関数である．
%続いて，より高精度な手法として，特異値に対する重み付け[細野ら]，および各特異値のp乗[Kongら]を行う二種類の手法が提案された．
%前者は行列の特異値に対する重み付き核ノルム[参考文献]の，後者は行列に対するSchatten-pノルム[参考]のテンソルに対する自然な拡張となっている．
%これらの手法は依然として連続関数でああり，どちらも非凸だが核ノルムより良い行列ランクの近似となっている．
%また，重み付けとp乗を組み合わせた手法も提案されている[zhangら]．

%3. 疑問は以下\\
%a) どっちが良いのか？組み合わせは効くのか？\\
%b) そもそもどっちも非凸なんだし，単純なランク最小化じゃだめ？\\
%Consequently, 以下のような自然な疑問がariseする．
Now, some natural questions arise:
%特異値に対する重み付けとp乗の効果は相乗的なのか？
%それともいずれかがいずれかを内包するのか？
Are the effects of the weightings for singular values and the Schatten-$p$ extension synergistic, or does one of them encompass?
%非凸関数を用いることを許すならば，単純なランク制約付き最小化の結果がこれらのadvancedな手法を上回ることはないのだろうか？
Is there any chance that a simple rank-constrained minimization, which is also a nonconvex optimization, can compete with these advanced and complicated methods? 

%4. 貢献は以下\\
%a) weight+Schatten-pの目的関数とアルゴリズムを提案．\\
%b) 網羅的な実験により，オラクルな重みが使えるならp=1でweightingするのが常に良いことを明らかにした．\\
%c) 重みが観測から得られる場合は組み合わせが有効な場合もある．\\
%以上の疑問に答えるために，まず本論文では低ランクテンソル復元のための，重み付けとp乗を組み合わせた一般的な新たな制約付き最小化問題と，それを解くアルゴリズムを提案した．
In this paper, to answer the questions above, we propose a novel general constrained optimization problem combining the weighting and the Schatten-$p$ extension for tensors, and we develop an efficient algorithm to solve it.
%網羅的な実験により，オラクル重みが使える場合において，p=1で重み付けすることが常に良いことを明らかにした．
We performed exhaustive experiments, and the results showed that if we can use the oracle weights, the combination of $p=1$ and the weighting is the most effective choice for all cases.
%また，観測信号から求めた重みを使う場合において，p=1/2と重み付けの組み合わせが効果を発揮することも確認できた．
We also found that the combination of $p=1/2$ and the weighting is effective when using the weights estimated from degraded measurements.
%ランク制約付き最小化問題では，制約として用いるランクを正しく決定できない限り，高品質な結果は得られないことが明らかとなった．
The rank-constrained minimization problem performs well as long as we know the rank of the original tensor. 
If we are agnostic toward the correct rank, the performance drops sharply.

%contribution一覧
%本論文における主な貢献を以下に示す．
The main contributions of this paper are summarized as follows:
\begin{itemize}
    %\item 特異値に対する重み付けとp乗の効果の間の関係性を明らかにするため，これらを組み合わせた一般的な新しい制約付き最小化問題と，これを解くアルゴリズムを提案した．
    \item  We propose a general constrained optimization problem and an efficient solver  for analyzing the relationship between the weightings of singular values and the Schatten-$p$ extension for tensors.
    %\item 重み付けとp乗は互いに包含関係になく，重みの決定方法によって最も性能の高くなるpが変化することを明らかにした．
    \item We show that the weighting and the Schatten-$p$ extension are synergetic and  that the effective value of $p$ is dependent on how the weights are determined.
    %\item 単純なランク制約付き最小化は制約に用いるランクに対して非常に敏感であり．元のテンソルのランクを正しく推定できない限り，提案するアルゴリズムのような高度な手法を上回ることはないことを明らかにした．
    \item We show that the rank constrained minimization problem is not able to outperform the advanced methods unless the true rank of the original tensor is known. The performance is sensitive to the rank values used as the constraints.
\end{itemize}

\section{Low-rank tensor completion}
\label{sec:pre}
In what follows, ${\mathbb N}$, $\R$ and $\R_+$ denote the set of all nonnegative integers, all real numbers, and all nonnegative real numbers.
We use capital calligraphic letters for tensors, capital bold letters for matrices, and lowercase bold letters for column vectors.

In this paper, we assume that an observation model of tensor recovery can be described as
\begin{equation}
    \label{eq:obsv}
    \cY = \mathrm{A}_\Omega(\cX_\mathrm{org}+\cV),
\end{equation}
where $\cY\in\R^{n_1\times\cdots\times n_N}$, $\cX_\mathrm{org}\in\R^{n_1\times\cdots\times n_N}$, and $\cV\in\R^{n_1\times\cdots\times n_N}$ are an $N$-th order observation tensor, an $N$-th order low-rank original tensor, and an $N$-th order random tensor, respectively, whose entries are i.i.d. Gaussian variables with zero mean and known variance $\sigma_n^2$.

The degradation operator is defined as
\begin{equation}
    \label{eq:degrad}
    \mathrm{A}_\Omega(\cX)_{i_1,\cdots,i_N} = \left\{ \begin{array}{ll}
    \cX_{i_1,\cdots,i_N} & ({i_1,\cdots, i_N}\in \Omega) \\
    0 & (\textrm{otherwise})
  \end{array} \right.
  ,
\end{equation}
where $\Omega$ is a set of indicators of observable entries.

%原テンソルが低ランクであると仮定するなら，これの推定は観測テンソルから近くかつランクの低いテンソルを求めることで実現できる．
If we can assume that the original tensor is low rank, it can be estimated by finding a tensor that is close to the observation tensor and also low rank.
%特に，原テンソルのランクが既知であり，観測テンソルにノイズが含まれていないと仮定すると，原テンソルの推定は，次のような原テンソルとランクが同じでかつ，観測テンソルと既知要素が一致するテンソル，すなわち下記の集合に含まれるテンソルをひとつ見つける問題となる．
In particular, assuming that the rank of the original tensor is known and that the variance of the noise is 0, estimating the original tensor is the problem of finding a tensor whose rank is identical to the rank of the original tensor and whose known elements match the observation tensor, i.e., finding a tensor within the following set:

\begin{equation}
\label{eq:rankeq+L2const}
    \textrm{Find}\ \cX\in\R^{n_1\times\cdots\times n_N}\ \textrm{s.t.}\ \mathrm{A}_\Omega(\cX)=\mathrm{A}_\Omega(\cY), \mathrm{rank}_m(\cX)=\hat{r}_m
\end{equation}
%ここで，$\mathrm{rank}_m(\cX)=\textrm{rank}(\unfold_m(\cX))$，$m=1,\cdots,N$，$\mathrm{rank}$は行列ランク，$\hat{r}_m$は原テンソルのmモード展開の行列ランクである．
where $\mathrm{rank}_m(\cX)=\textrm{rank}(\unfold_m(\cX))$ and $m=1,\cdots,N$. 
We denote $\mathrm{rank}$ as the matrix rank and $\hat{r}_m$ as the matrix rank of an $m$-th mode unfolded original tensor.

%しかしながら，式（\ref{eq:rankeq+L2const}）のようにランクに関する等式を含む集合は扱いづらい．
However, in general, the set containing the equations for the ranks, as in Eq. (\ref{eq:rankeq+L2const}), is hard to determine.
%そこで式(3)のかわりに，次のような不等式制約を用いた集合へと変更する．
Thus, we employ an alternative set with an inequality constraint instead of Eq.  (\ref{eq:rankeq+L2const}):
\begin{equation}
\label{eq:rankineq+L2const}
    \textrm{Find}\ \cX\in\R^{n_1\times\cdots\times n_N}\ \textrm{s.t.}\ \mathrm{A}_\Omega(\cX)=\mathrm{A}_\Omega(\cY), \mathrm{rank}_m(\cX)\leq\hat{r}_m,
\end{equation}

%式（\ref{eq:rankeq+L2const}），(\ref{eq:rankineq+L2const})のような集合は，一般に複数の要素を持つためこれをそのまま原テンソルの推定に用いるのは難しい．
The sets shown in Eqs. (\ref{eq:rankeq+L2const}) and (\ref{eq:rankineq+L2const}) are used directly for the estimation of the original tensor because they generally contain multiple matrices.
%また，観測過程にノイズが含まれている場合には空集合となる可能性がある．
Additionally, if the observation process includes noise ($\sigma_N \neq 0$), it may yield the empty set.
%そこで，ガウス分布の負の対数尤度と対応している観測テンソルとの間でのL2ノルムを用いて，次のような最小化問題の解を推定テンソルとする．
Thus, we employ the L2 norm between the observation tensor, which corresponds to the negative log-likelihood of the Gaussian distribution, and the solution of the following minimization problem, including this norm as the estimated tensor.
\begin{equation}
    \label{eq:rankinneq+L2fid}
    \min_{\cX\in\R^{n_1\times\cdots\times n_N}} \|\A_\Omega(\cX)-\cY\|_2^2\ \ 
    \textrm{s.t.}\ \mathrm{rank}_m(\cX)\leq\hat{r}_m (m=1,\cdots.N),
\end{equation}
%ここで，$\|\cdot\|_2$ はテンソルの$\ell_2$ノルムであり，テンソルの各要素の二乗の和の平方根である．
where $\|\cdot\|_2$ is an $\ell_2$ norm of the tensor that is defined as the square root of the sum of the squares of each element of the tensor.
%残念ながらこの問題は非凸最適化だが，Alternating Direction Method of Multipliers（ADMM）\cite{Gabay1976}を用いて解くことができる．
%ADMMは凸最適化問題を解くためのアルゴリズムだが，非凸最適化問題に対しても実用上有効であることが知られている\cite{Chartrand2013,Sun2014,Ono2017}．
Although that this problem is one of nonconvex optimization, we can efficiently solve it by using the alternating direction method of multipliers (ADMM) \cite{Gabay1976}, which is known as  an algorithm for solving convex optimization problems and is effective in practice for solving nonconvex optimization problems \cite{Chartrand2013,Sun2014,Ono2017}.

%式（\ref{eq:rankeq+L2const}）--（\ref{eq:rankinneq+L2fid}）では各undoldしたテンソルの行列ランクを見ているが，行列ランクは非凸なだけでなく不連続であり，非常に扱いづらい．
Eqs. (\ref{eq:rankeq+L2const}) and (\ref{eq:rankinneq+L2fid}) include the matrix rank of each unfolded tensor, which is very difficult to handle since it is not only nonconvex but also discontinuous.
%また，原テンソルの（unfoldingした各行列の）ランク($\hat{r}_m$)が，事前に正確にわかっていることも稀である．
Moreover, the situation in which we know the rank of each unfolded matrix of the original tensor  $\hat{r}_m$ is unrealistic.

%そこで，非凸だが連続なテンソルランクの表現として，Weighted tensor schatten-$p$ norm が提案されている．
The weighted tensor Schatten-$p$ norm (WTSPN) is proposed as a representation of a nonconvex but continuous tensor rank, 
\begin{equation}
    \label{eq:wtspn}
    \|\cdot\|_{\w, \bvec{\gamma}, p}\colon\R^{n_1\times\cdots\times n_N}\to\R_+\colon\cX\mapsto\sum_{m=1}^N\gamma_m\|\unfold_m(\cX)\|_{\w_m,p}^p,
\end{equation}
where $\|\cdot\|_{\w,p}^p$ is a weighted Schatten-$p$ norm raised to the power $p$ (WSPN)  \cite{Zha2018}; $\w=[\w_1^\top, \cdots, \w_N^\top]^\top$ is the weight vector of the WSPNs; $\gamma_m$ is a positive constant satisfying $\sum_{m=1}^N\gamma_m=1$;  $\bvec{\gamma}=[\gamma_1, \cdots, \gamma_N]^\top$, and $\unfold(\cdot)$ is an unfolding operator. 

The WTSPN is generally a nonconvex function that is consistent with the weighted tensor nuclear norm \cite{Hosono2019} when $p=1$, the tensor Schatten-$p$ norm \cite{Kong2018} when $\w$ is uniform  (all elements of $\w$ are the same value), and the tensor nuclear norm \cite{Gandy2011, Liu2013} when $p=1$ and $\w$ is uniform.

The $m$-th mode tensor unfolding operator of the $n$-th order tensor $\unfold_m:\R^{n_1\times\cdots\times n_N}\to\R^{n_m\times I_m}$ is defined as a map from the tensor elements ($i_1,\cdots,i_N$) to the corresponding matrix elements ($i_m,j_m$), where $I_m = \prod^N_{\substack{k=1\\k\neq m}}i_k$, 

\begin{equation}
j_m=1+\sum^N_{\substack{k=1\\k\neq m}}(i_k-1)\prod^{k-1}_{\substack{l=1\\l\neq m}}i_l.
\end{equation}

The WSPN is described as
\begin{equation}
\label{eq:wspn}
\|\cdot\|_{\w,p}^p\colon\R^{n_v\times n_h}\to\R_+\colon\X\mapsto\sum_{k=1}^{n_m}w_k\sigma_k(\X)^p,
\end{equation}
where $0<p$, $n_m=\textrm{min}(n_v,n_h)$, 
$\sigma_k(\X)\in\R_+\ (k=1,\cdots,n_m)$ is the $k$-th largest singular value of $\X$, 
and $\w=[w_1,\cdots,w_{n_m}]^{\top}\in\R^{n_m}_+$ is a weight vector that satisfies $0\leq w_1\leq w_2\leq \cdots \leq w_{n_m} $.
The WSPN is a generalization of the nuclear norm and the weighted nuclear norm \cite{Chen2013,Gu2017}, which are often used in low-rank matrix recovery.

%\ref{sec:intro}章で述べたように，WTSPNの適切な重みや$p$の決定に関しては詳細な調査がされてきておらず，それらを明らかにするのが本研究の目的のひとつである．
As mentioned in Section \ref{sec:intro}, the proper weights of the WTSPN and the proper value of  $p$ have not been investigated in detail. 
Revealing them is one of the objectives of this paper.
%ところで，WTSPNを正則化項，L2ノルムを忠実化項として式（\ref{eq:obsv}）の観測モデルから復元を行う場合，ノイズ分散が変わらなくても正則化項のパラメータ$\w$や$p$によって，最適なハイパーパラメータ（正則化項と忠実加工のバランスを取るパラメータなど）が変化してしまい，公正な比較が難しくなる．

On the other hand, when recovering the observation model of Eq. (\ref{eq:obsv}) with the WTSPN as the regularization term and the $\ell_2$ norm as the fidelity term, even if the noise variance does not change, the optimal hyperparameter (balancing the regularization term and the fidelity term) varies according to the parameters of the regularization term, $\w$, and $p$.
This makes a fair comparison difficult.
%また，実用上もこのようなチューニングの難しいパラメータが存在するのは望ましくない．
In addition, a parameter that is so difficult to tune is not desirable for practical use.
%そこで，次節でこの問題を解決するための方法を提案する．
Therefore, in the next section, we propose a method to solve this problem.

\section{Proposed method}
%前述の問題を解決するため，L2ボール制約とWTSPN最小化を用いる方法を提案する．
To solve the above problem, we propose a method using $\ell_2$ ball constraints and WTSPN minimization.
%具体的には，次のような最小化問題を用いる
Specifically, we formulate the following minimization problem
\begin{equation}
    \label{eq:prop}
    \min_\cX \|\cX\|_{\w,\bvec{\gamma},p}\ \ \ \mathrm{s.t.}\
    \mathrm{A}_\Omega(\cX)\in{\mathrm{B}(\cY,\sigma_n\sqrt{|\Omega|})},
\end{equation}
%ここで，$\mathrm{B}(\y,r)$はL2ボールであり，センター$\y\in\R^N$ ，半径$r\in\R$のボールは次のように定義される
where $\mathrm{B}(\cY,r)$ is an $\ell_2$ ball, and the $\ell_2$ ball with center $\cY\in\R^{n_1\times\cdots\times n_N}$ and radius $r\in\R$ is defined as
\begin{equation}
    \label{eq:ball}
    \mathrm{B}(\cY,r):=\{\cX\in\R^{n_1\times\cdots\times n_N} | \|\cX-\cY\|_2\leq r\}.
\end{equation}
%ボール制約を用いることで，ノイズの分散のみから適切なパラメータを決定することが可能となり，本論文のように様々な正則化パラメータを比較する際に都合が良い．
By using the ball constraint, it is possible to determine the appropriate parameters based only on the variance of the noise\cite{Afonso2011,Chierchia2015,Ono2015}, which is convenient when comparing various regularization parameters, as in this paper.
%また，$|\Omega|$は集合$\Omega$の要素数である．
Additionally, $|\Omega|$ is the number of elements of the set $\Omega$.

%我々はノイズの標準偏差$\sigma_n$を既知と仮定しているため，その標準偏差によって定まる超球の内側に，原テンソルに対して付加されたノイズ$\cV$の実現値が存在することが期待できる．
Since we assume that the standard deviation of noise $\sigma_n$ is known, we can expect the realization of the noise $\cV$ added to the original tensor to exist inside the hypersphere determined by the standard deviation.
%式（\ref{eq:prop}）の制約はこの事実と対応している．
The constraints of Eq. (\ref{eq:prop}) accord with this fact.
%この工夫は我々に，「（ノイズの分散が既知とするなら）異なる正則化項の能力を公平に比較すること」を可能とする．
This method allows us to ``fairly compare the performance of different regularization terms (if the variance of the noise is known).''

%この問題は一般に非凸最適化問題となり，大域的最適解を見つけるのは困難である．
In general, Eq. (\ref{eq:prop}) is a nonconvex optimization problem, which makes it difficult to find a globally optimal solution.
As mentioned in Section \ref{sec:pre}, ADMM exhibits empirical performance on nonconvex optimization problems.
%そこで我々は交互方向乗数法（ADMM）を用いて式（\ref{eq:prop}）を解くことを提案する．
Therefore, we propose solving Eq. (\ref{eq:prop}) using ADMM.
%提案するアルゴリズムをAlgorithm \ref{al:ADMM}に示す．
The proposed algorithm is shown in Algorithm \ref{al:ADMM}.
%ここでも我々はADMMを用いて非凸最適化の問題を解いている．

%Algorithm \ref{al:ADMM}の5行目
The objective function in line 5 of Algorithm \ref{al:ADMM} is 
\begin{equation}
    \label{eq:pro1}
    \argmin_\X \|\X\|_{\lambda\gamma_m\w_m,p}^p+\frac{1}{2}\|\unfold_m(\cX^{(k+1)})+\Z_{1,m}^{(k)}-\X\|_F^2
\end{equation}
%の目的関数は非凸だが，解のうちの一つは次のように書ける\cite{Xie2016,Zhang2019,Zha2018}.
which is nonconvex, although one of the solutions can be written as \cite{Xie2016,Zhang2019,Zha2018}:
\begin{equation}
    \label{eq:proxofwsp}
    \U\mathrm{S}_{\lambda\gamma_m\w_m,p}(\Sigma)\V^\top,
\end{equation}
%ここで，$\U\Sigma\V^\top=\unfold_m(\cX^{(k+1)})+\Z_{1,m}^{(k)}$は特異値分解，$\mathrm{S}_{\w,p}(\cdot)$ は重み付きthresholding operatorである．
where $\U\Sigma\V^\top=\unfold_m(\cX^{(k+1)})+\Z_{1,m}^{(k)}$ is a singular value decomposition (SVD) and $\mathrm{S}_{\w,p}(\cdot)$ is a weighted thresholding operator.

%対角行列のみに値を持つ行列$Y$に対するthresholding operatorの各要素$(\mathrm{S}_{\w,p}(\Y))_{i,i}$は最小化問題
Each element of the weighted thresholding operator for a rectangular diagonal matrix $Y$ $(\mathrm{S}_{\w,p}(\Y))_{i,i}$ is defined as a solution to the following minimization problem:
\begin{equation}
    \label{eq:lpreg}
    \argmin_\x \frac{1}{2}(\Y_{i,i}-\x)^2+\w_i|\x|^p.
\end{equation}
%の解として定義される．
%式（\ref{eq:lpreg}）の解は$p=1$のときsoft thresholding $\max(\Y_{i,i}-\w_i,0)$であり，$p=\{1/2,2/3\}$のとき文献\cite{Cao2013}で提案されたclosed form thresholdingによって解が得られる．
The solution of Eq. (\ref{eq:lpreg}) is a soft thresholding $\max(\Y_{i,i}-\w_i,0)$ when $p=1$ and the closed-form thresholding proposed in \cite{Cao2013} when $p=\{1/2,2/3\}$.

%Algorithm \ref{al:ADMM}の8行目の第1項
The first term in line 8 of Algorithm \ref{al:ADMM} is
\begin{equation}
    \label{eq:pro2}
    \proj_{\mathrm{B}(\cY,\sigma_n\sqrt{|\Omega|})}(\mathrm{A}_{\Omega}(\cX^{(k+1)}+\cZ_{2}^{(k)}))
\end{equation}
%は集合$\mathrm{B}(\cY,\sigma_n\sqrt{|\Omega|})$に対する距離射影
which is a metric projection of the set $\mathrm{B}(\cY,\sigma_n\sqrt{|\Omega|})$.
The metric projection is defined as
\begin{equation}
    \label{eq:proj}
    \textrm{proj}_S \colon\R^N\to\R^N\colon\x\mapsto\argmin_{\y\in S}\frac{1}{2}\|\x-\y\|^2_2.
\end{equation}

%であり，式（\ref{eq:pro2}）は次のように書ける
Eq. (\ref{eq:pro2}) has a closed-form solution,
\begin{equation}
    \label{eq:solveproj}
    %c-min(r/norm(y-c,2),1)*(c-y)
    \cY-\min\left(\frac{\sigma_n\sqrt{|\Omega|}}{\|\cY-\mathrm{A}_{\Omega}(\cX^{(k+1)}+\cZ_{2}^{(k)})\|_2},1\right)(\cY-\mathrm{A}_{\Omega}(\cX^{(k+1)}+\cZ_{2}^{(k)})).
\end{equation}

%また，8行目の第2項
The set of $\bar{\Omega}$ in the second term of line 8,
\begin{equation}
    \mathrm{A}_{\bar{\Omega}}(\cX^{(k+1)}+\cZ_{2}^{(k)})
\end{equation}
%の$\bar{\Omega}$はobserving entry indicator $\Omega$の補集合であり，indicators of missing entriesの集合である．
is the complement of the set $\Omega$, which is a set of indicators of missing entries.
%$\mathrm{A}_{\bar{\Omega}}$は以下のように定義される
$\mathrm{A}_{\bar{\Omega}}$ is defined as
\begin{equation}
    \label{eq:degradcmp}
    \mathrm{A}_{\bar{\Omega}}(\cX)_{i_1,\cdots,i_N} = \left\{ \begin{array}{ll}
    \cX_{i_1,\cdots,i_N} & ({i_1,\cdots, i_N}\in \bar{\Omega}) \\
    0 & (\textrm{otherwise})
  \end{array} \right.
  .
\end{equation}

\begin{algorithm}[t]
\caption{Proposed algorithm}
\label{al:ADMM}
{%\small
\begin{algorithmic}[1]
\REQUIRE $\cY$, $\sigma_n$, $\bvec{\gamma}=[\gamma_1,\cdots,\gamma_N]$, $\w=[\w_1^\top,\cdots,\w_N^\top]^\top$, $p$, $\lambda$
\STATE{\textrm{Initialize} $\Y_{1,m}^{(0)}=\unfold_m(\cY)$, $\cY^{(0)}_2=\cY$, $\Z_{1,m}^{(0)}=0$, $\Y_{1,m}^{(0)}$, $\cZ_2^{(0)}=0$}
\WHILE{\textit{A stopping criterion is not satisfied}}
	\STATE $\cX^{(k+1)}=\argmin_{\cX}\frac{1}{2}\sum_{m=1}^N\|\Y_{1,m}^{(k)}-\unfold_m(\cX)-\Z_{1,m}^{(k)}\|^2_2$\\
	$+\lambda\|\cY_2^{(k)}-\mathrm{A}_\Omega(\cX)-\cZ_2^{(k)}\|^2_2$
	\FOR{$m = 1$ to $N$}
	    \STATE $\Y_{1,m}^{(k+1)}=\argmin_\X \|\X\|_{\lambda\gamma_m\w_m,p}^p$\\
	    $+\frac{1}{2}\|\unfold_m(\cX^{(k+1)})+\Z_{1,m}^{(k)}-\X\|_F^2$
	    \STATE $\Z_{1,m}^{(k+1)}=\Z_{1,m}^{(k)}+\unfold_m(\cX^{(k+1)})-\Y_{1,m}^{(k+1)}$
	\ENDFOR
	\STATE $\cY_2^{(k+1)}=\proj_{\mathrm{B}(\cY,\sigma_n\sqrt{|\omega|})}(\mathrm{A}_{\Omega}(\cX^{(k+1)}+\cZ_{2}^{(k)}))+\mathrm{A}_{\bar{\Omega}}(\cX^{(k+1)}+\cZ_{2}^{(k)})$
	\STATE $\cZ_2^{(k+1)}=\cZ_2^{(k)}+\mathrm{A}_{\Omega}(\cX^{(k+1)})-\cY_2^{(k+1)}$
	\STATE $\lambda = 0.99\lambda$
	\STATE $k=k+1$
\ENDWHILE
\ENSURE $\cX^{(k)}$
\end{algorithmic}
}
\end{algorithm}

\section{Experimental Comparison}

\def\order{3}
\def\rank{4}

\def\mpsize{8.5}
\begin{figure*}
\begin{center}
\centering
%\begin{tabular}{cc}
%---------------------
\vspace{0.5cm} 
\def\mr{40}
\def\nl{0}
\begin{minipage}[t]{\mpsize cm}
\begin{center}
\includegraphics[width=\linewidth]{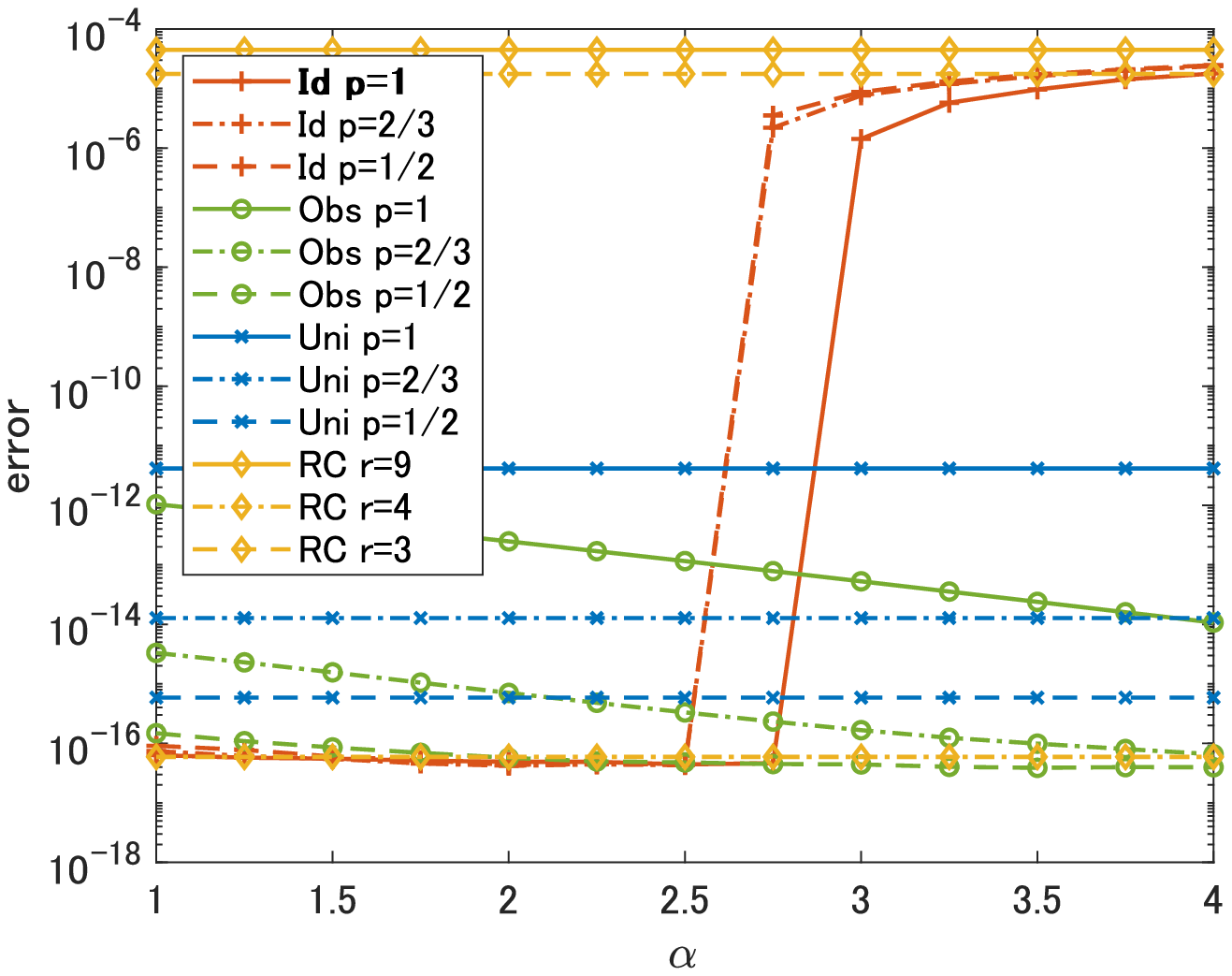}
\subcaption{missing rate \mr\% $\sigma_n=\nl$ \\}
\end{center}
\end{minipage}
%---------------------
\vspace{0.5cm}
\def\mr{80}
\def\nl{0}
\begin{minipage}[t]{\mpsize cm}
\begin{center}
\includegraphics[width=\linewidth]{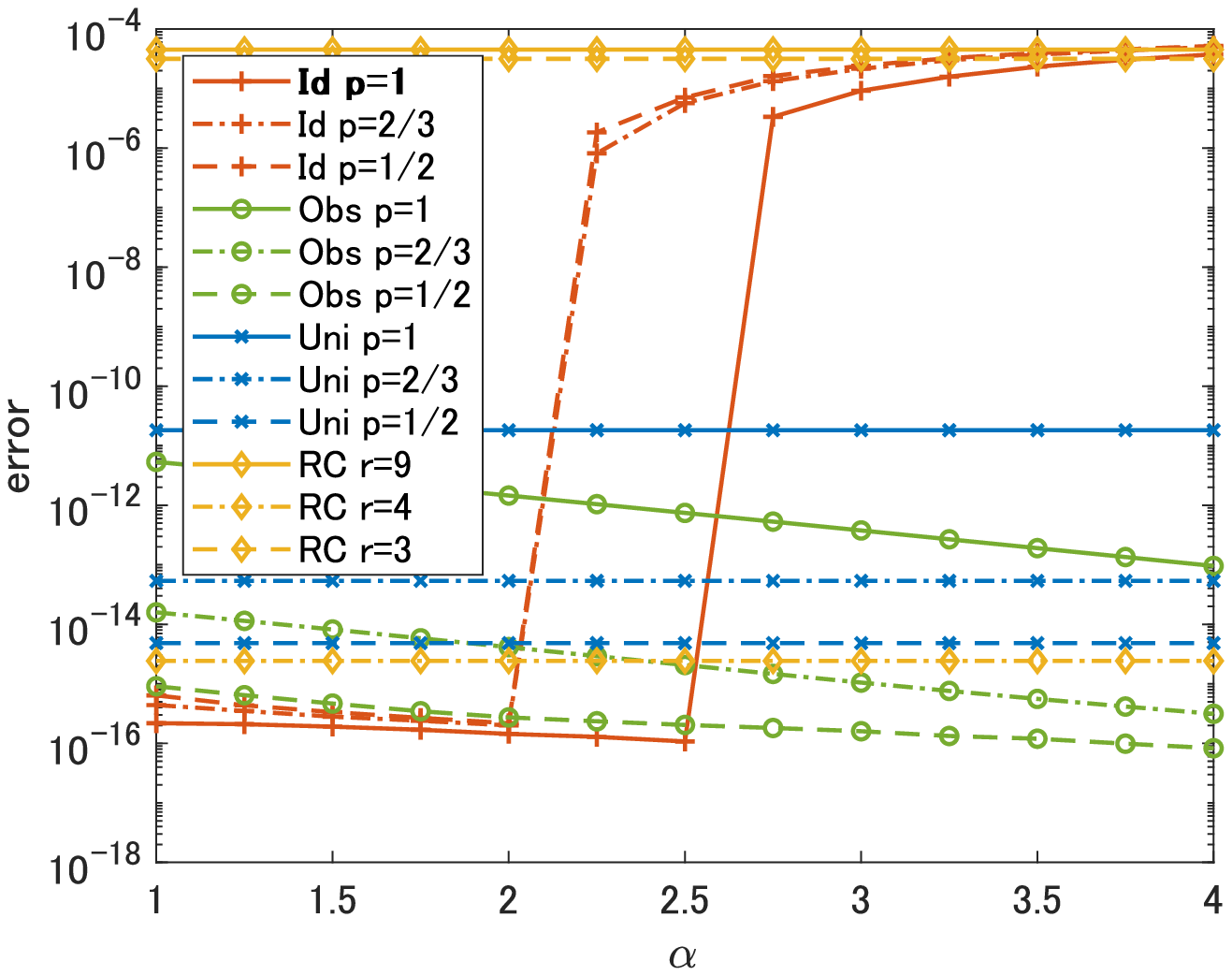}
\subcaption{missing rate \mr\% $\sigma_n=\nl$ \\}
\end{center}
\end{minipage}
%---------------------
\vspace{0.5cm}
\def\mr{40}
\def\nl{1}
\begin{minipage}[t]{\mpsize cm}
\begin{center}
\includegraphics[width=\linewidth]{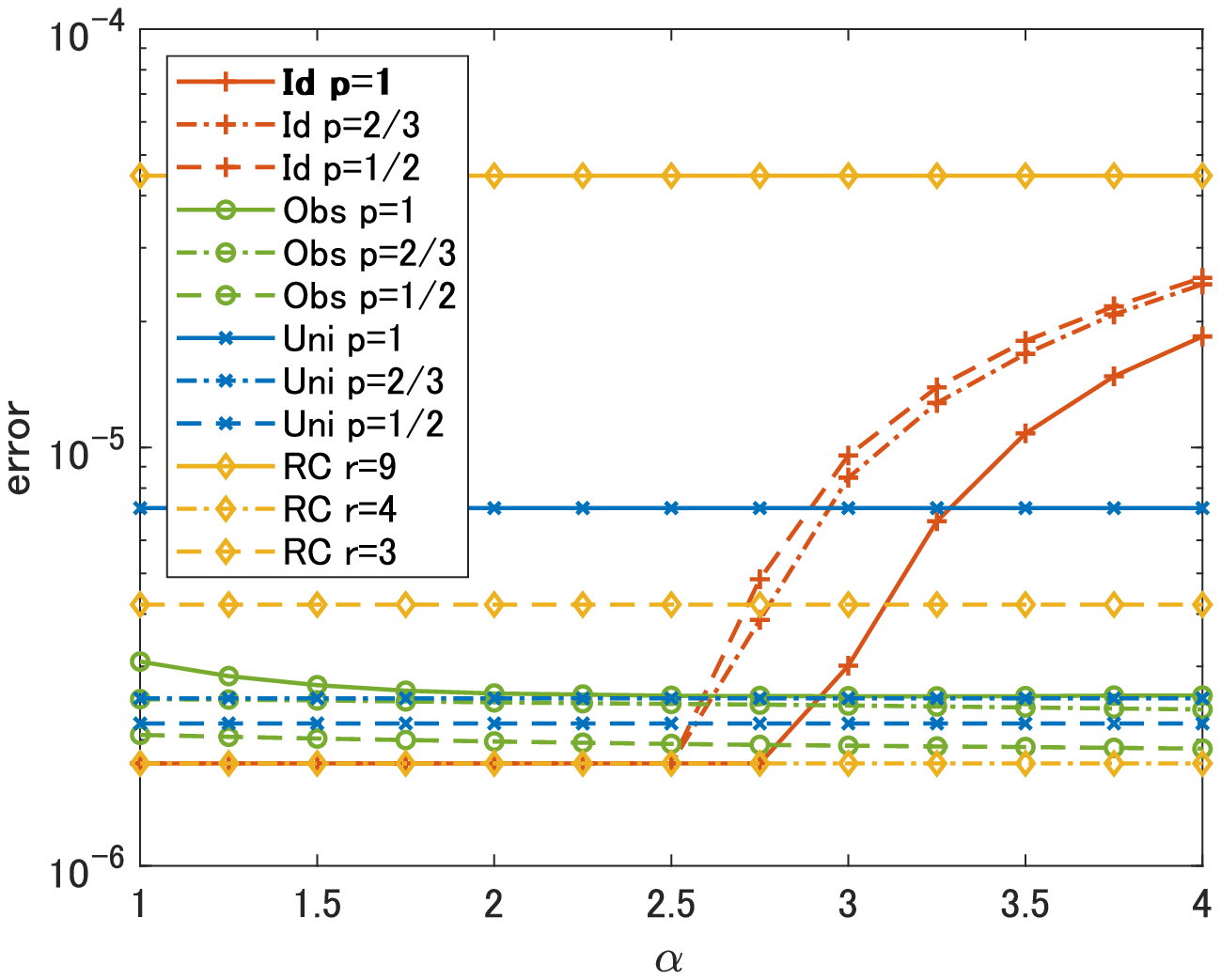}
\subcaption{missing rate \mr\% $\sigma_n=\nl$}
\end{center}
\end{minipage}
%---------------------
\vspace{0.5cm}
\def\mr{80}
\def\nl{1}
\begin{minipage}[t]{\mpsize cm}
\begin{center}
\includegraphics[width=\linewidth]{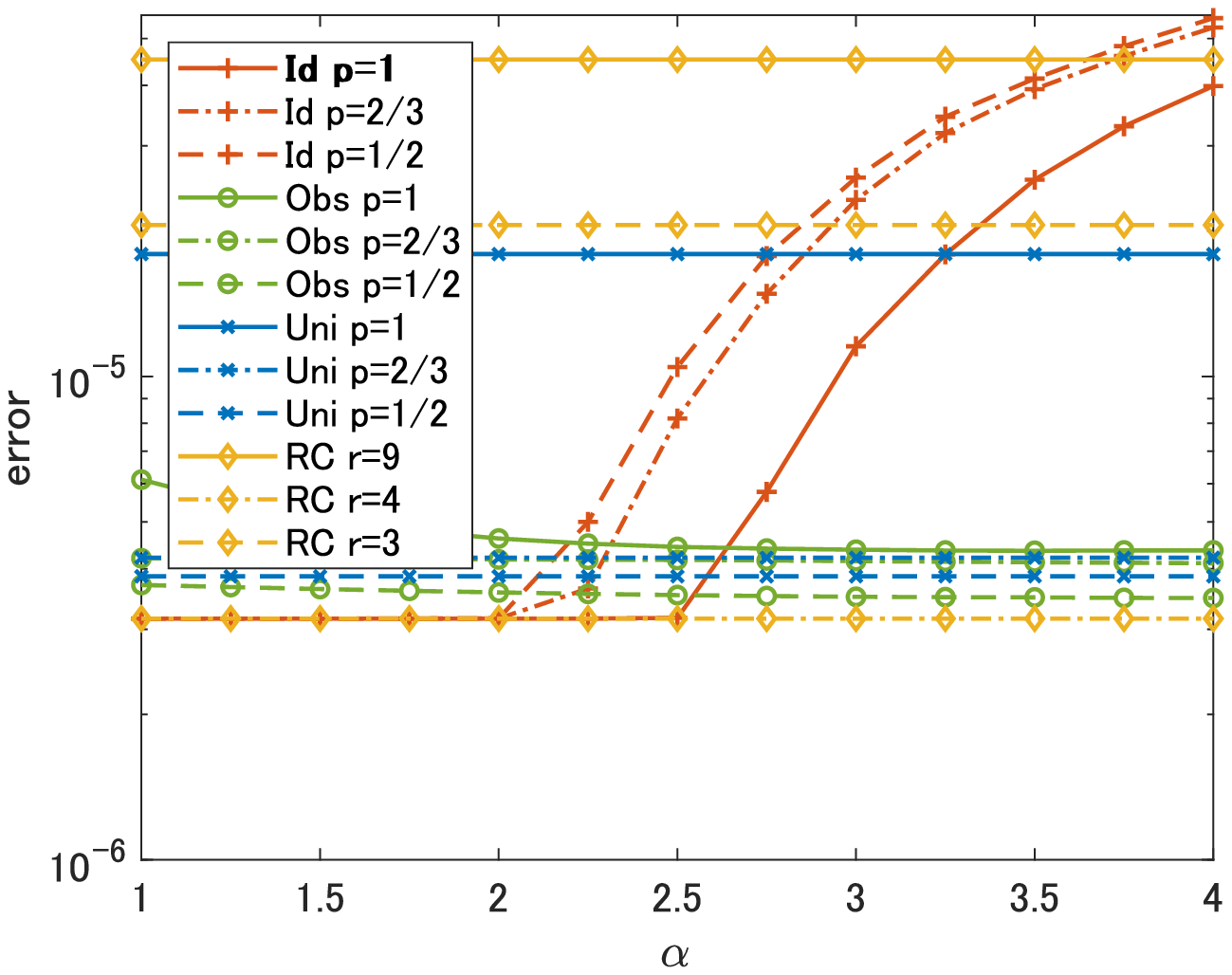}
\subcaption{missing rate \mr\% $\sigma_n=\nl$ \\}
\end{center}
\end{minipage}
\caption{%原テンソルを階数\order，ランク\rankとしたときの実験結果．
Experimental results for the original tensor with the order \order \ and the rank \rank.
%(a)-(d)はそれぞれ観測過程での欠損率とノイズの標準偏差$\sigma_n$を変化させたときの結果である．
(a)-(d) are the results of varying the missing rate and the standard deviation of the noise $\sigma_n$ during the observation process.
%グラフの横軸は重み決定のためのパラメータ$\alpha$，縦軸は各手法により求めた推定テンソルと原テンソルの間の誤差である．
The horizontal axis of the graph is the parameter $\alpha$ used for determining the weights and the vertical axis is the error between the estimated tensor and the original tensor calculated by each method.
%赤，緑，青がそれぞれ提案手法(Algorithm\ref{al:ADMM})で異なる種類の重みベクトル$\w_{Id}$,$\w_{Obs}$, $\w_{Uni}$を用いたときの結果を，黄はランク制約付き最小化（Algorithm ?（appendに入れる予定のAlg）)のときの結果である．線種はパラメータ$p$の値と対応している．
Red, green and blue indicate the results for the proposed method (Algorithm \ref{al:ADMM}) with different types of weight vectors--$\w_\mathrm{Id}$,$\w_\mathrm{Obs}$, and  $\w_\mathrm{Uni}$, respectively--and yellow is the result of the rank-constrained minimization.
The line type corresponds to the value of the parameter $p$ or $r$.
\label{fig:o3r4}}
%\end{tabular}
\end{center}
\end{figure*}

%\newpage
\def\order{3}
\def\rank{5}

\def\mpsize{8.5}
\begin{figure*}
\begin{center}
\centering
%---------------------
\vspace{0.5cm}
\def\mr{40}
\def\nl{0}
\begin{minipage}[t]{\mpsize cm}
\begin{center}
\includegraphics[width=\linewidth]{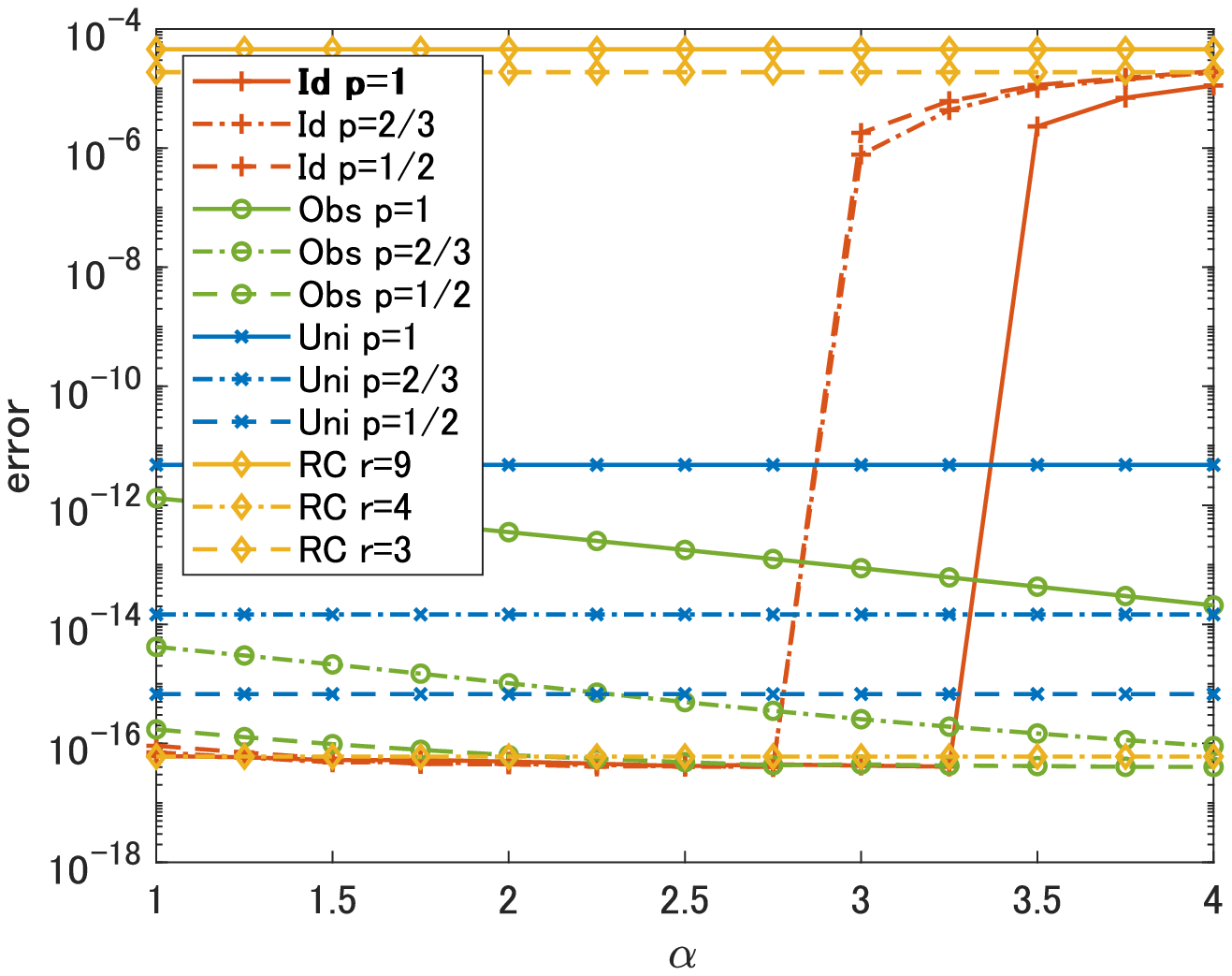}
\subcaption{missing rate\mr\% $\sigma_n=\nl$ \\}
\end{center}
\end{minipage}
%---------------------
\vspace{0.5cm}
\def\mr{80}
\def\nl{0}
\begin{minipage}[t]{\mpsize cm}
\begin{center}
\includegraphics[width=\linewidth]{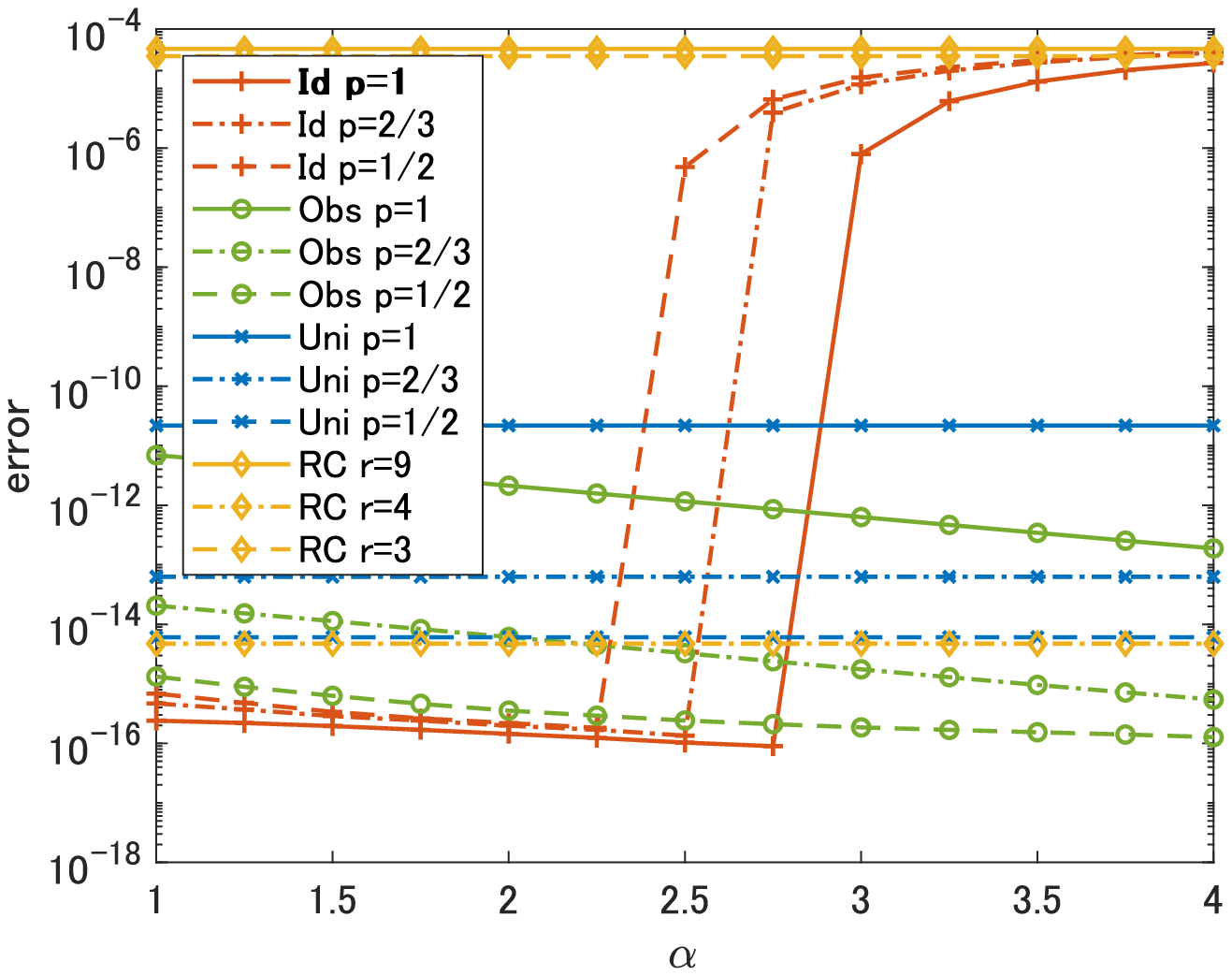}
\subcaption{missing rate\mr\% $\sigma_n=\nl$ \\}
\end{center}
\end{minipage}
%---------------------
\vspace{0.5cm}
\def\mr{40}
\def\nl{1}
\begin{minipage}[t]{\mpsize cm}
\begin{center}
\includegraphics[width=\linewidth]{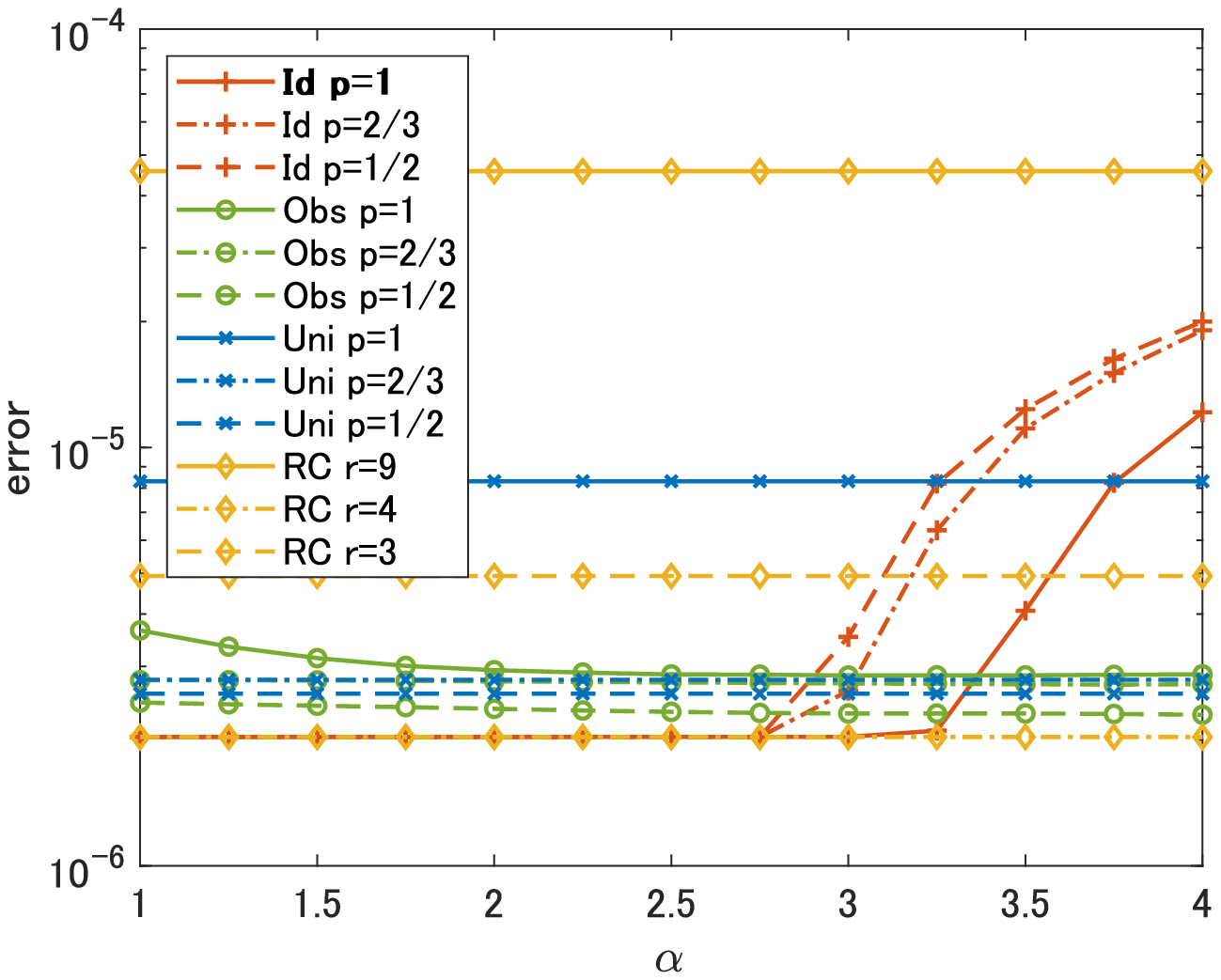}
\subcaption{missing rate\mr\% $\sigma_n=\nl$}
\end{center}
\end{minipage}
%---------------------
\vspace{0.5cm}
\def\mr{80}
\def\nl{1}
\begin{minipage}[t]{\mpsize cm}
\begin{center}
\includegraphics[width=\linewidth]{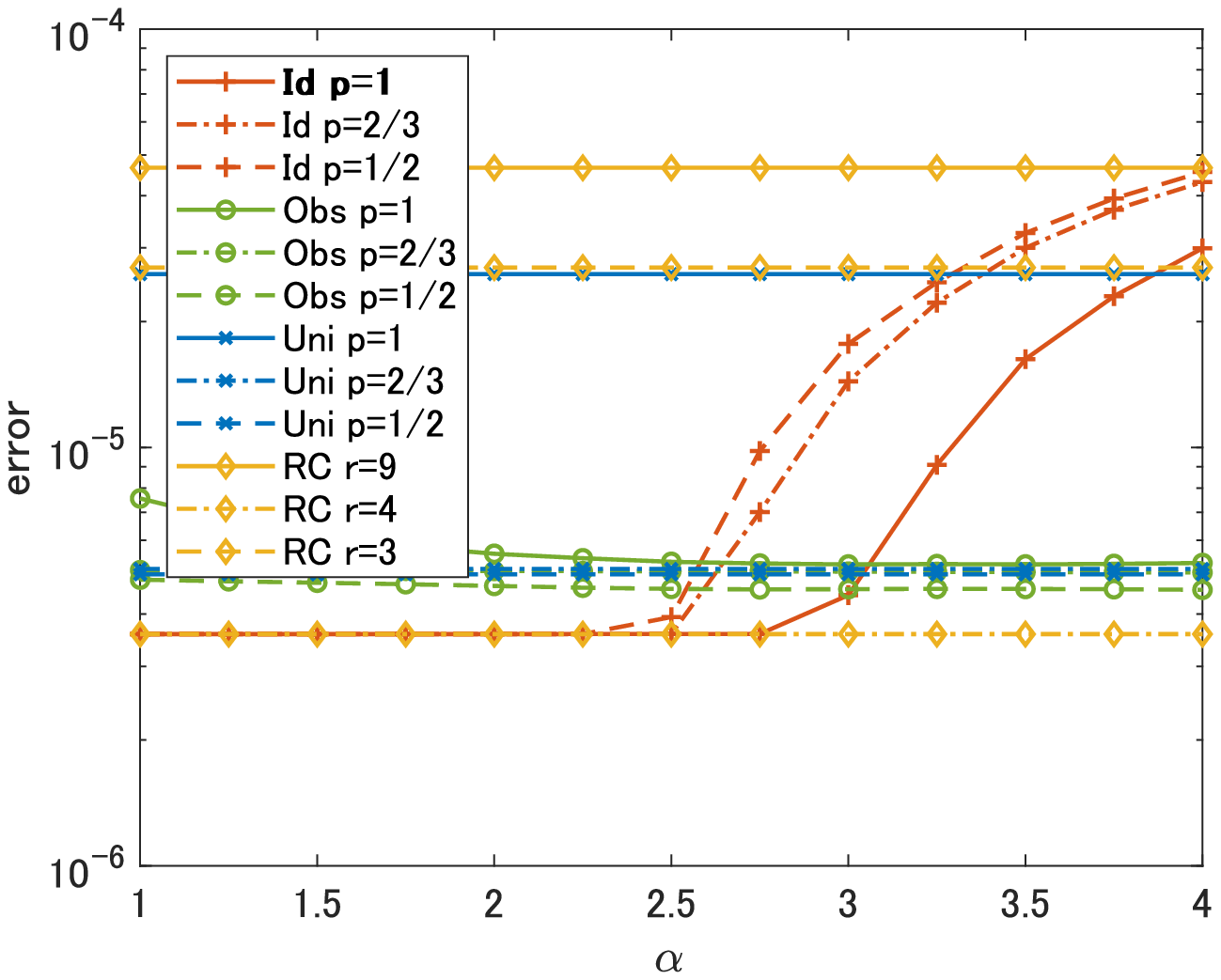}
\subcaption{missing rate\mr\% $\sigma_n=\nl$ \\}
\end{center}
\end{minipage}
\caption{%原テンソルを階数\order，ランク\rankとしたときの実験結果．　
%(a)-(d)はそれぞれ観測過程での欠損率とノイズの標準偏差$\sigma_n$を変化させたときの結果である．
%グラフの横軸は重み決定のためのパラメータ$\alpha$，縦軸は各手法により求めた推定テンソルと原テンソルの間の誤差である．
Experimental results for the original tensor with the order \order \ and the rank \rank.
The results of different missing rates and the standard deviations are shown in (a)-(d) and the meaning of the axes and of the colors and types of lines are the same as in Fig. \ref{fig:o3r4}.
%Fig.\ref{fig:o3r4}と比較すると原テンソルのランクが異なるが，各手法間の相対的な性能に関しては同様の傾向が見られる．
Although the rank of the original tensor is different, the relative performance of all methods shows a similar trend in Fig. \ref{fig:o3r4}.
This supports the fact that our conclusions in section \ref{subsec:discussion} are independent of the rank of the original tensor.\label{fig:o3r5}}
\end{center}
\end{figure*}

%\newpage
\def\order{4}
\def\rank{2}

\def\mpsize{8.5}
\begin{figure*}
\begin{center}
\centering
%---------------------
\vspace{0.5cm}
\def\mr{40}
\def\nl{0}
\begin{minipage}[t]{\mpsize cm}
\begin{center}
\includegraphics[width=\linewidth]{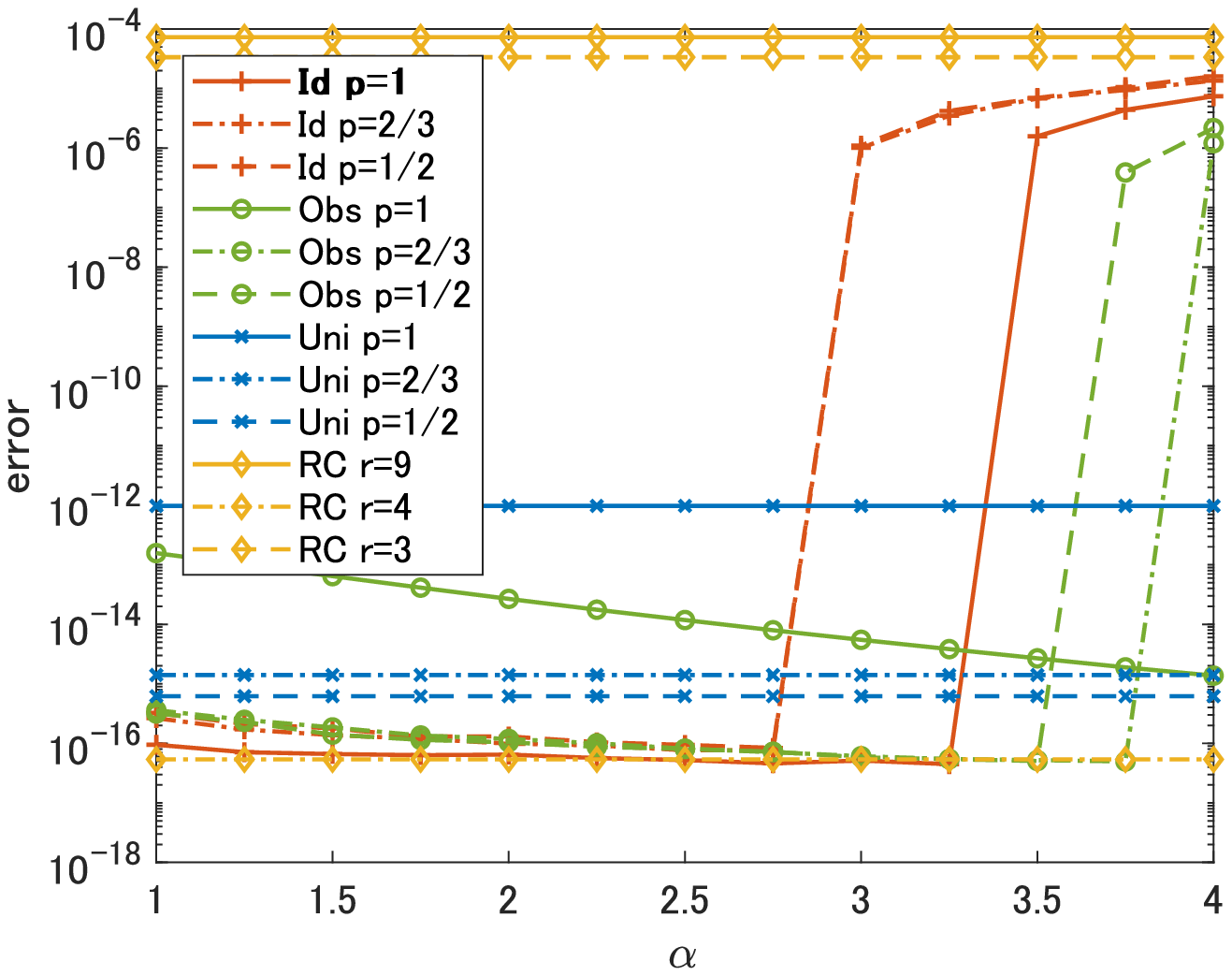}
\subcaption{missing rate\mr\% $\sigma_n=\nl$ \\}
\end{center}
\end{minipage}
%---------------------
\vspace{0.5cm}
\def\mr{80}
\def\nl{0}
\begin{minipage}[t]{\mpsize cm}
\begin{center}
\includegraphics[width=\linewidth]{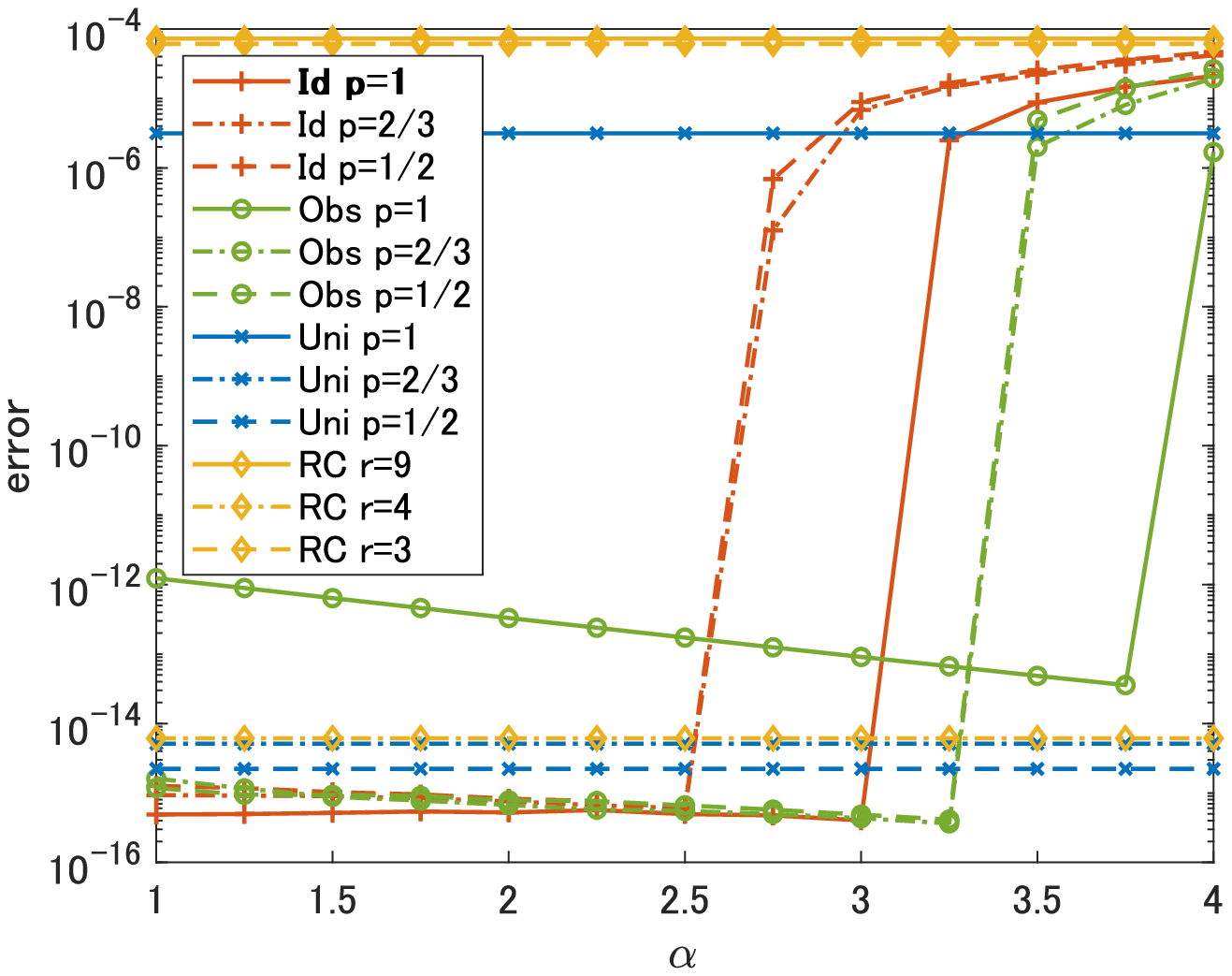}
\subcaption{missing rate\mr\% $\sigma_n=\nl$ \\}
\end{center}
\end{minipage}
%---------------------
\vspace{0.5cm}
\def\mr{40}
\def\nl{1}
\begin{minipage}[t]{\mpsize cm}
\begin{center}
\includegraphics[width=\linewidth]{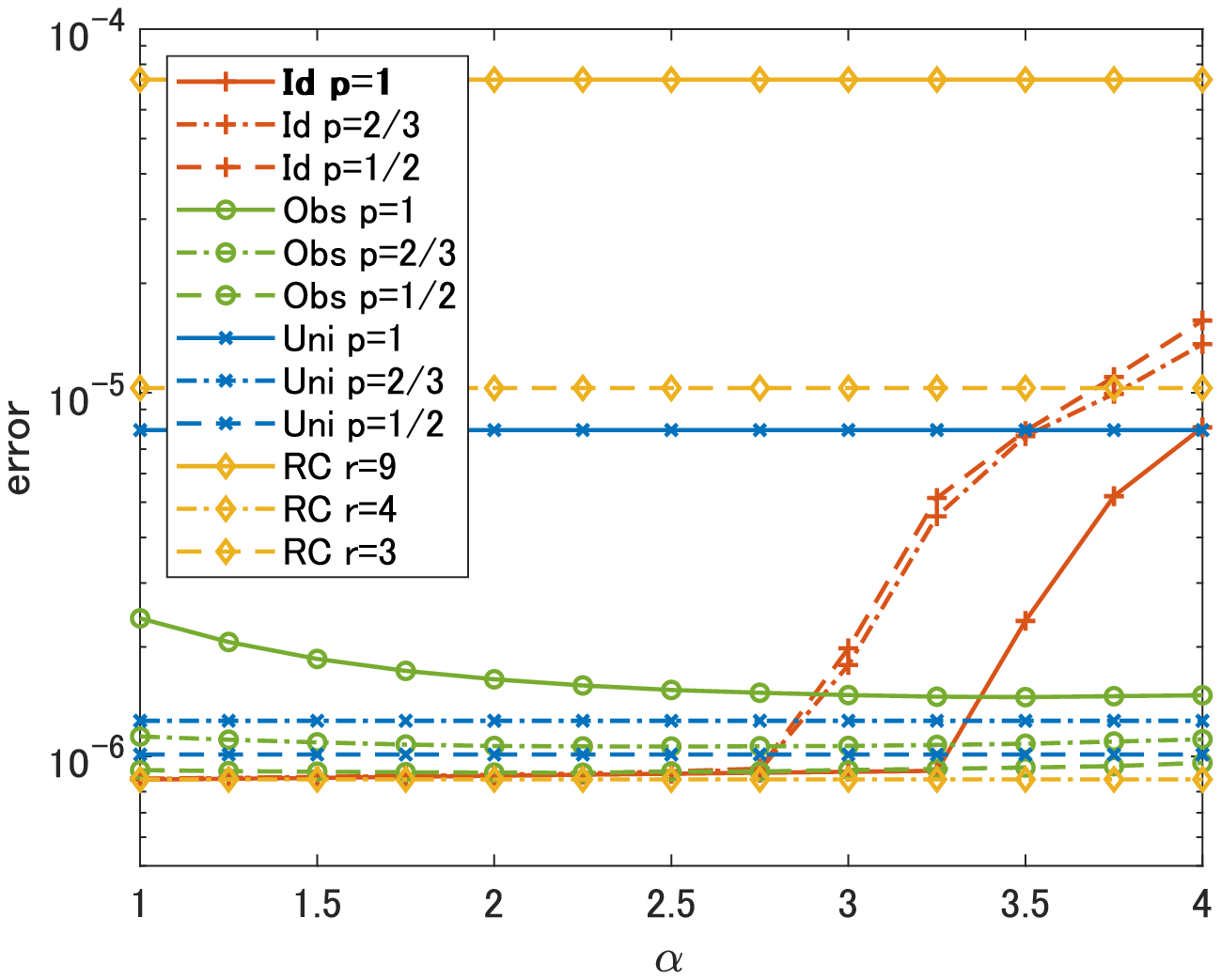}
\subcaption{missing rate\mr\% $\sigma_n=\nl$}
\end{center}
\end{minipage}
%---------------------
\vspace{0.5cm}
\def\mr{80}
\def\nl{1}
\begin{minipage}[t]{\mpsize cm}
\begin{center}
\includegraphics[width=\linewidth]{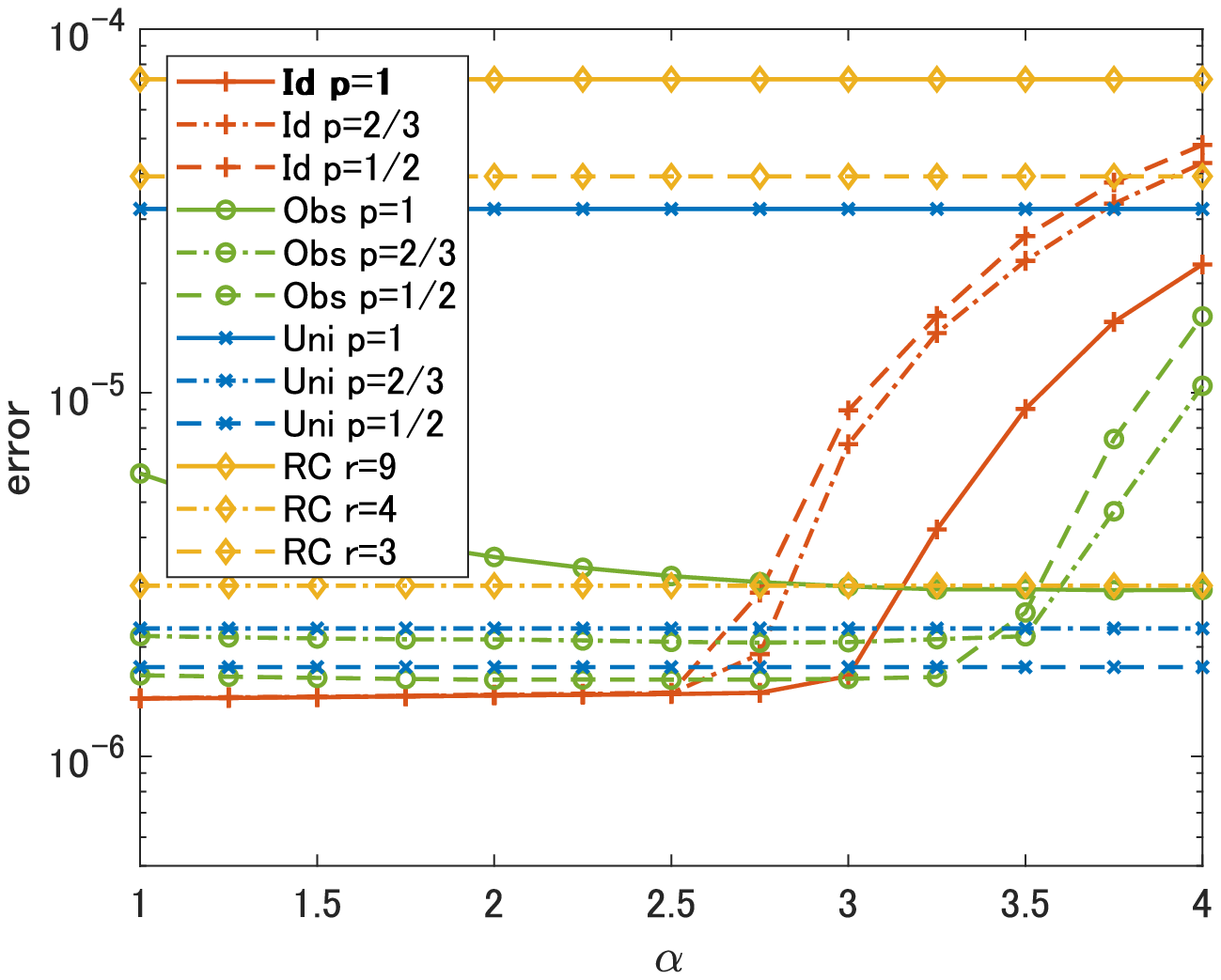}
\subcaption{missing rate\mr\% $\sigma_n=\nl$ \\}
\end{center}
\end{minipage}
\caption{%原テンソルを階数\order，ランク\rankとしたときの実験結果．　
%(a)-(d)はそれぞれ観測過程での欠損率とノイズの標準偏差$\sigma_n$を変化させたときの結果である．
%グラフの横軸は重み決定のためのパラメータ$\alpha$，縦軸は各手法により求めた推定テンソルと原テンソルの間の誤差である．
%各線の色と種類の意味はFig.\ref{fig:o3r4}と同様である．
Experimental results for the original tensor with the order \order \ and the rank \rank.
The results of different missing rates and the standard deviations are shown in (a)-(d) and the meaning of the axes and of the colors and types of lines are the same as in Fig. \ref{fig:o3r4}.
%Fig.\ref{fig:o3r4}と比較すると原テンソルの階数，ランクが異なるが，各手法間の相対的な性能に関しては同様の傾向が見られる
Although the rank and the order of the original tensor are different, the relative performance of all methods shows a similar trend to Fig. \ref{fig:o3r4}.
This supports the fact that our conclusions in section \ref{subsec:discussion} are independent of the rank and the order of the original tensor.\label{fig:o4r2}}
\end{center}
\end{figure*}

%\newpage
\def\order{4}
\def\rank{3}

\def\mpsize{8.5}
\begin{figure*}
\begin{center}
\centering
%---------------------
\vspace{0.5cm}
\def\mr{40}
\def\nl{0}
\begin{minipage}[t]{\mpsize cm}
\begin{center}
\includegraphics[width=\linewidth]{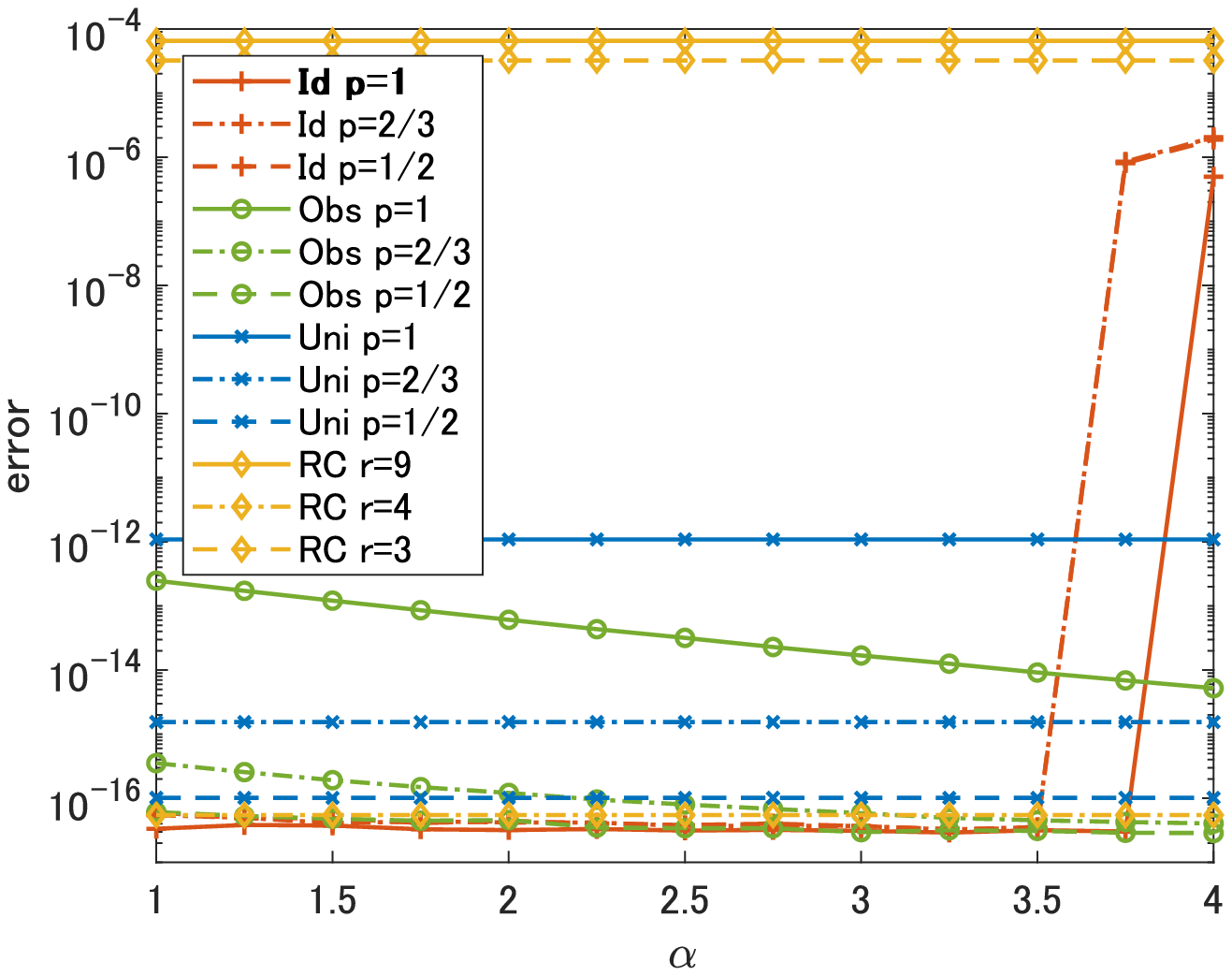}
\subcaption{missing rate\mr\% $\sigma_n=\nl$ \\}
\end{center}
\end{minipage}
%---------------------
\vspace{0.5cm}
\def\mr{80}
\def\nl{0}
\begin{minipage}[t]{\mpsize cm}
\begin{center}
\includegraphics[width=\linewidth]{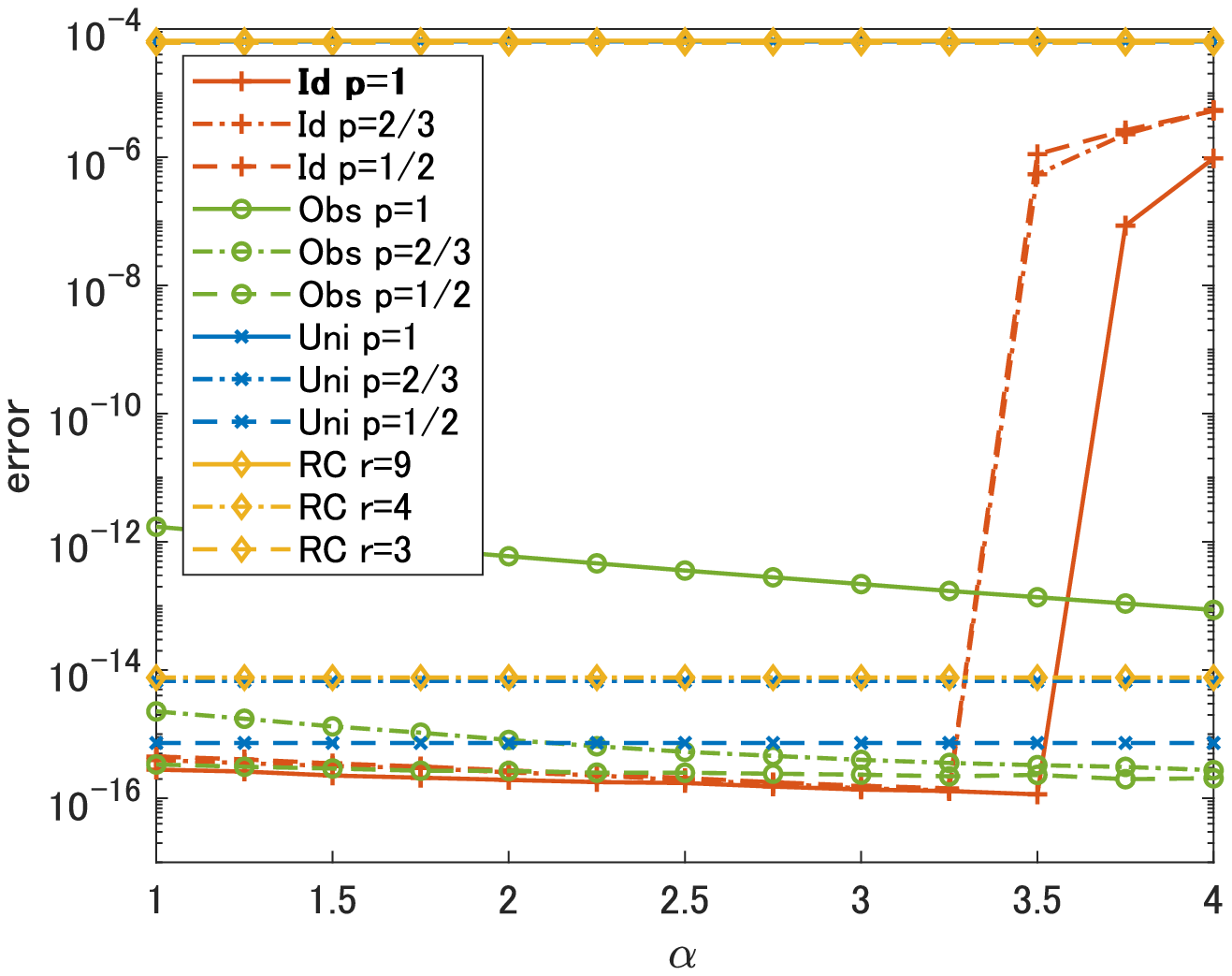}
\subcaption{missing rate\mr\% $\sigma_n=\nl$ \\}
\end{center}
\end{minipage}
%---------------------
\vspace{0.5cm}
\def\mr{40}
\def\nl{1}
\begin{minipage}[t]{\mpsize cm}
\begin{center}
\includegraphics[width=\linewidth]{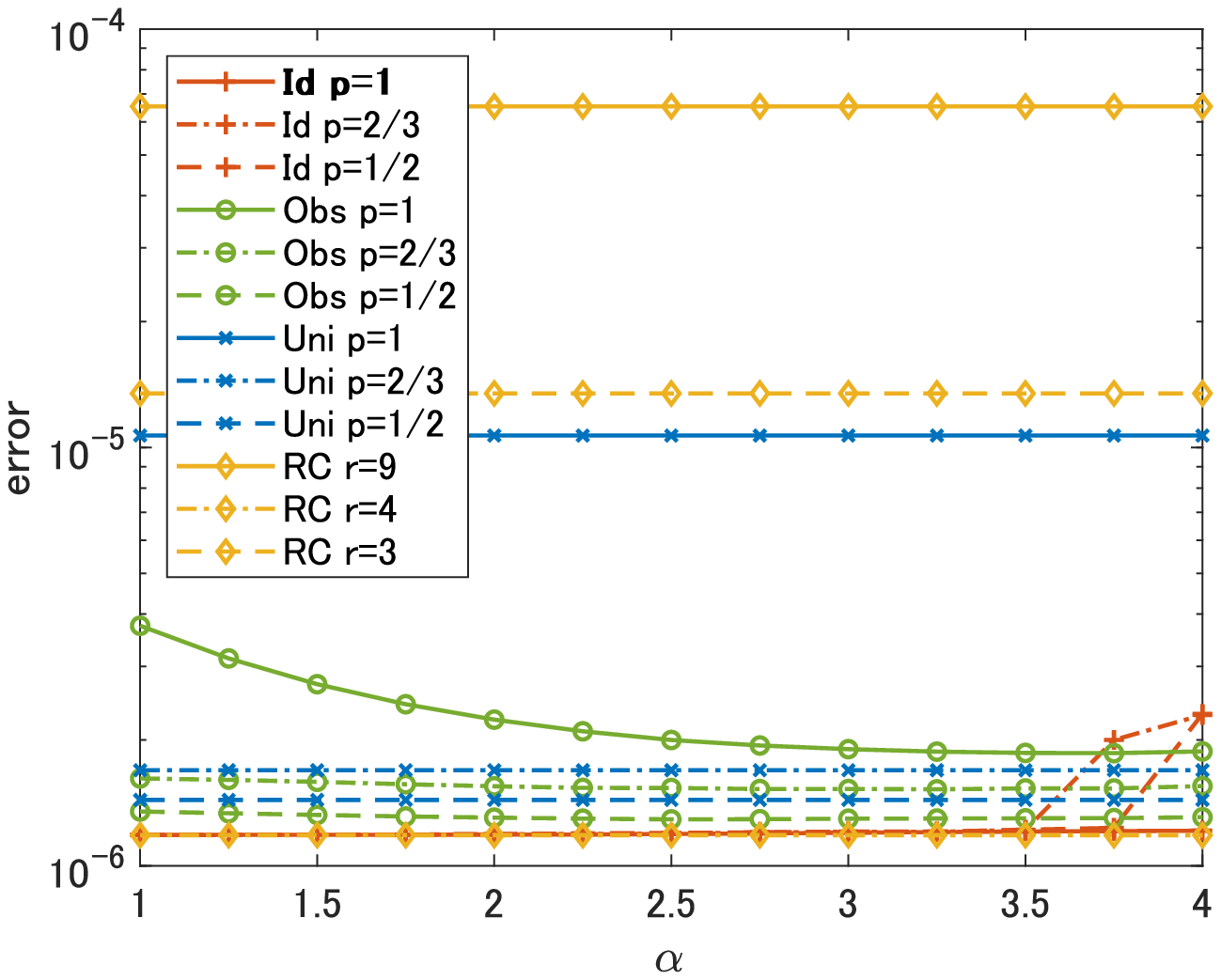}
\subcaption{missing rate\mr\% $\sigma_n=\nl$}
\end{center}
\end{minipage}
%---------------------
\vspace{0.5cm}
\def\mr{80}
\def\nl{1}
\begin{minipage}[t]{\mpsize cm}
\begin{center}
\includegraphics[width=\linewidth]{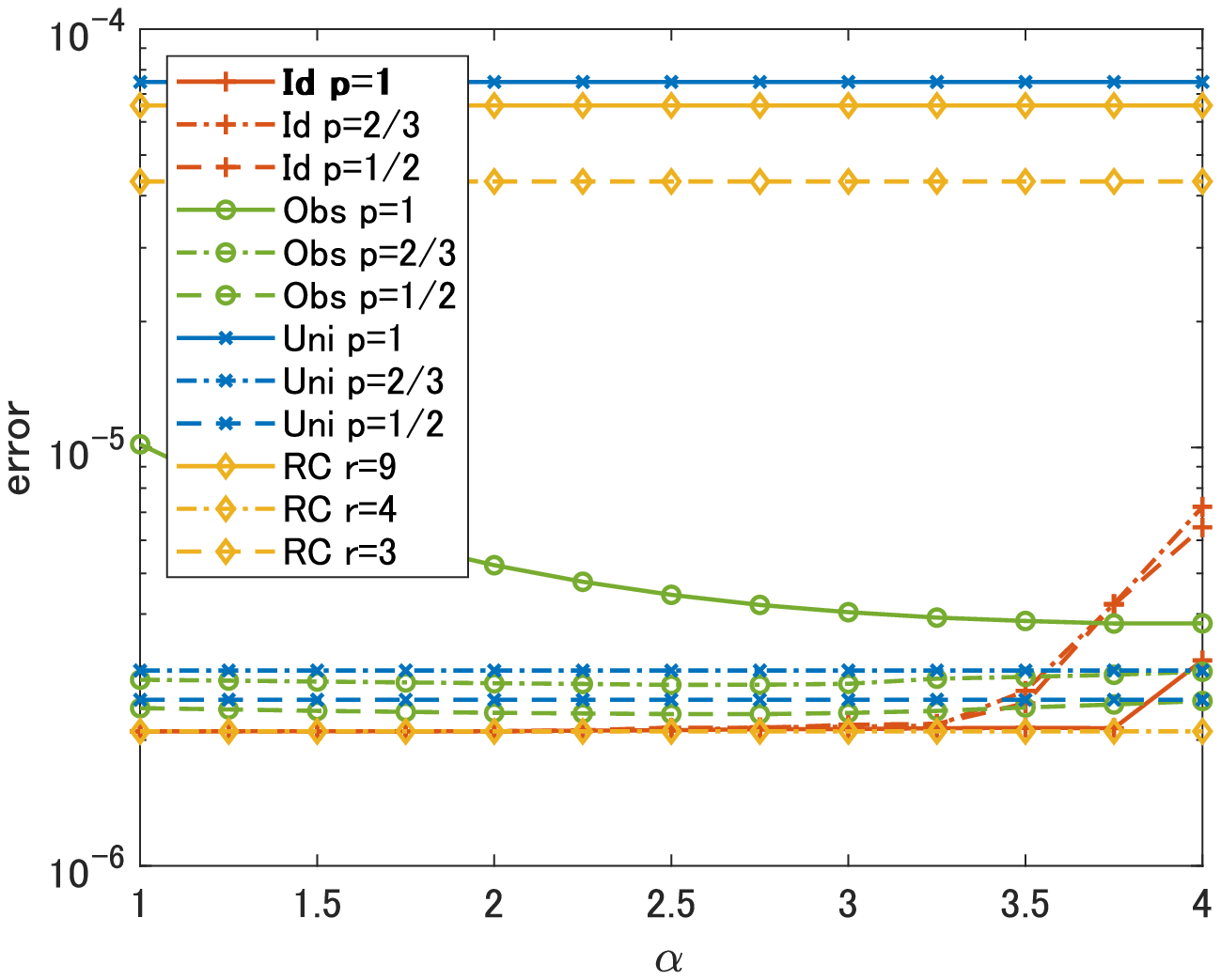}
\subcaption{missing rate\mr\% $\sigma_n=\nl$ \\}
\end{center}
\end{minipage}
\caption{%原テンソルを階数\order，ランク\rankとしたときの実験結果．　
%(a)-(d)はそれぞれ観測過程での欠損率とノイズの標準偏差$\sigma_n$を変化させたときの結果である．
%グラフの横軸は重み決定のためのパラメータ$\alpha$，縦軸は各手法により求めた推定テンソルと原テンソルの間の誤差である．
%各線の色と種類の意味はFig.\ref{fig:o3r4}と同様である．
Experimental results for the original tensor with the order \order \ and the rank \rank.
The results of different missing rates and the standard deviations are shown in (a)-(d) and the meaning of the axes and of the colors and types of lines are the same as in Fig. \ref{fig:o3r4}.
%Fig.\ref{fig:o3r4}と比較すると原テンソルの階数，ランクが異なり，Fig.\ref{fig:o4r2}とは原テンソルのランクが異なるが，各手法間の相対的な性能に関しては同様の傾向が見られる
Although the rank and the order of the original tensor are different, the relative performance of all methods shows a similar trend to Fig. \ref{fig:o3r4}.
This supports the fact that our conclusions in section \ref{subsec:discussion} are independent of the rank and the order of the original tensor.\label{fig:o4r3}}
\end{center}
\end{figure*}

\subsection{Setting}
\label{subsec:setting}
%導入
%Introductionで提示した2つの疑問，特異値に対する重み付けとp乗の効果は相乗的なのか？それとも一方がもう一方を包含するのか？，単純なrank制約付き最小化では不十分なのか？に答えるため，人工テンソルを用いた実験を行った．
In section \ref{sec:intro}, we posed two questions:
\begin{itemize}
    \item Are the effects of the weighting and $p$-squared on singular values synergistic? Or does one encompass the other?
    \item Is simple rank-constrained minimization insufficient? 
\end{itemize}
To answer these questions, we performed some experiments using an artificial tensor.
%テンソルの作り方
%$N$階人工テンソル$\cX\in\R^{n_1\times\cdots\times n_N}$の各要素は，タッカーモデルを用いて次のように生成する．
Each element of an $N$-th order artificial tensor $\cX\in\R^{n_1\times\cdots\times n_N}$ is generated by using the Tucker model:
\begin{equation}
    \cX(i_1,\cdots,i_N) = \sum_{1\leq j_1\leq r_1,\cdots,1\leq j_N\leq r_N} \cS(j_1,\cdots,j_N)\prod_k^N \U_k(i_k,j_k),
\end{equation}
%ここで，$[r_1,\cdots,r_N]$の各要素は$\cX$を各モードでunfoldしたときの行列ランク，$\cS\in\R^{r_1\times\cdots\times r_N}$はコアテンソル，$\U_k\in\R^{n_k\times r_k}$はfactorテンソルである．
where $[r_1,\cdots,r_N]$, $\cS\in\R^{r_1\times\cdots\times r_N}$ and $\U_k\in\R^{n_k\times r_k}$ are the matrix ranks of $\unfold_m(\cX)$\ $(m = 1,\cdots,N)$, the core tensor, and the factor tensor.
%$\cS$と$\U_k$の各要素はそれぞれ区間$[0,1]$の一様分布乱数と区間$[-0.5,0.5]$の一様分布乱数を用いて決定した．
Each element of $\cS$ and $\U_k$ is generated uniformly over the intervals $[0,1]$ and $[-0.5,0.5]$, respectively.
%最後に，$\cX$の要素の最大値と最小値の差分が$1$になるよう正規化を行った．
Finally, we normalized the difference between the maximum and minimum elements of $\cX$ to $1$.

%手法，重みの決め方
%重み付き核ノルムでよく用いられる重みとして，原行列特異値の逆数（オラクル重み）がある．
%Section 1で述べたように，特異値に対する理想的な重みとしてよく挙げられるのは，原テンソルをunfoldしたものの特異値の逆数（オラクル重み）であり，次のように定義される．
As mentioned in section \ref{sec:intro}, the widely used weights used as the ideal weights for the singular values are the inverses of the singular values of the unfolded original tensor $\cX_\mathrm{org}$.
Throughout this paper, we define our \textit{ideal weights} as 
\begin{equation}
\label{eq:wid}
    (\w_{\mathrm{Id}}(\alpha))_{i,j} = R\frac{\sigma_i(\unfold_j(\cX_\mathrm{org}))^{-\alpha}}{\sum_k\sigma_k(\unfold_j(\cX_\mathrm{org}))^{-\alpha}},
\end{equation}
%ここで，$R$は$\unfold_j(\cX)$の行数，列数のうち小さい方である．
where $R$ is smaller of the row and column dimensions of $\unfold_j(\cX_\mathrm{org})$.
%また，理想的な重みとされているものが復元性能の意味で最適とは限らないため，パラメータ$\alpha$を導入することでより柔軟に重みを設定することを可能としている．
Since the ideal weights are not always optimal in terms of the recovery performance, we introduce a parameter $\alpha$ to bring additional flexibility to the setting of the weights.
%しかしながら，実用上真の特異値を得られるケースは稀である．
However, the true singular values are not available in practical applications.

%そこで用いられるのが，観測から真の特異値を何らかの方法で推定し，それを用いて決定される重み（観測重み）であり，そのうちの一つとして次のようなものが考えられる．
A method that does not require the true singular values is to use the singular values obtained from observations for estimating the ideal weights. 
One of these methods is as follows:
\begin{equation}
\label{eq:wde}
    (\w_{\mathrm{Obs}}(\alpha))_{i,j} = R\frac{\sigma_i(\unfold_j(\tilde{\cY}))^{-\alpha}}{\sum_k\sigma_k(\unfold_j(\tilde{\cY}))^{-\alpha}}
\end{equation}
%ここで，$\tilde{\cY}$は観測テンソル$\cY$の欠損要素を非欠損要素の平均値で埋めたテンソルである．
where $\tilde{\cY}$ is a tensor the missing elements of the observation tensor $\cY$ filled in using the average of the observed entries of the observation tensor $\cY$.
%二種類の重みを用いて実験を行うことで，理想的なケースと現実的なケースの両方でweightingとSchatten-pの関係を明らかにすることができる．
We refer to these weights as \textit{observation weights}.

%また，重みを用いず$p$乗のみを行うtensor Schatten-$p$ normは，WTSPNの重みベクトルが全て同一の要素を持つ特別な場合となる．
The tensor Schatten-$p$ norm with no weights is a special case of the WTSPN.
%よって，tensor Schatten-$p$ normとして，次のような特別な重みを使ったWTSPNを用いる．
Therefore, we can use the WTSPN with the following special weights as the tensor Schatten-$p$ norm:
\begin{equation}
\label{eq:wun}
    (\w_{\mathrm{Uni}})_{i,j} = 1.
\end{equation}

In the following experiments, we use three types of weights (the ideal weights $\w_\textrm{Id}$, observation weights $\w_\textrm{Obs}$ and uniform weights $\w_\textrm{Uni}$) to reveal the relationship between the weighting and Schatten-$p$ extension on the performance of WTSPN.
%重み決定のためのパラメータ$\alpha$は$[1,4]$の範囲を$0.25$刻みで変化させた．
The weight determination parameter $\alpha$ varies in increments of $0.25$ in the range  $[1,4]$.
%Algorithm \ref{al:ADMM}のパラメータは$p=\{1/2, 2/3, 1\}$，$\bvec{\gamma}$の各要素は対象とするテンソルの階数$N$を用いて$1/N$，$\lambda=100$とした．
The parameters of Schatten-$p$ are chosen from $p=\{1/2, 2/3, 1\}$, and each element of $\bvec{\gamma}$ is set to $1/N$, where $N$ is the order of the target tensor.
In all cases, the parameter of ADMM is set to $\lambda=100$．
%lambdaの値はGithubに上がってるコードで確認したが最終commitが結構前だと思うので要確認←たぶん平気

%オラクル重み$\w_\mathrm{Id}$を用いたAlgorithm \ref{al:ADMM}，観測重み$\w_\mathrm{Obs}$を用いたAlgorithm \ref{al:ADMM}，均一重み$\w_\mathrm{Uni}$を用いたAlgorithm \ref{al:ADMM}，ランク制約付き最小化の性能を比較した．
We compare the performance of Algorithm \ref{al:ADMM} using $\w_\mathrm{Id}$, $\w_\mathrm{Obs}$, and $\w_\mathrm{Uni}$ as well as rank-constrained minimization.
%また，各手法の性能評価には以下のような誤差を用いた．
The following error is used to evaluate the performance of each method:
\begin{equation}
\label{eq:error}
    \textrm{error}(\tilde{\cX},\cY) = \frac{1}{\prod_{m=1}^N I_m}\|\tilde{\cX}-\cY\|_2,
\end{equation}
%ここで，$\tilde{\cX}$は各手法で得られた推定テンソルである．
where $\tilde{\cX}$ is the estimated tensor obtained by each method.
%rse=|X-X^*|_2/N HaLRTCはrerative squared error

\subsection{Results and Discussion}
\label{subsec:discussion}
%階数，ランク，欠損率，ノイズレベルに記号振って初回以外それで書く←やっぱ書かない
%欠損率を0.4と0.8の二種類，ノイズの標準偏差$\sigma_n$を0と1の二種類とに設定した観測テンソルに対して復元を行い，それらの四種類の組み合わせを各Figureの(a)から(d)に示した．
We performed recovery from observed tensors with missing rate of $0.4$ and $0.8$ and standard deviations of noise $\sigma_n$ of $0$ and $1$, and the results of the combinations of these  parameters are shown in (a) to (d) in Figs. \ref{fig:o3r4}, \ref{fig:o3r5}, \ref{fig:o4r2}, and \ref{fig:o4r3}.
%グラフの横軸は式（\ref{eq:wid}），（\ref{eq:wde}）で用いられている重み決定のためのパラメータ$\alpha$，縦軸は式（\ref{eq:error}）で定義される各手法の性能である．
The horizontal axis of each graph is the parameter $\alpha$ used in the Eqs. (\ref{eq:wid}) and (\ref{eq:wde}) for determining the weights, and the vertical axis is the performance of each method defined by Eq. (\ref{eq:error}).
%各線の色はそれぞれ赤が$\w_\mathrm{Id}$を用いたAlgorithm \ref{al:ADMM}（prop $\w_\mathrm{Id}$），緑が$\w_\mathrm{Obs}$を用いたAlgorithm \ref{al:ADMM}（prop $\w_\mathrm{Obs}$），青が$\w_\mathrm{Uni}$を用いたAlgorithm \ref{al:ADMM}（prop $\w_\mathrm{Uni}$），黄がランク制約付き最小化（rank const）となっている．
The red, green, blue, and yellow lines show the results of Algorithm \ref{al:ADMM} with $\w_\mathrm{Id}$ (Id in the legends of the graphs), with $\w_\mathrm{Obs}$ (Obs),  with $\w_\mathrm{Uni}$ (Uni), and with the rank-constrained minimization shown in Eq. (\ref{eq:rankinneq+L2fid}) (RC).
%また，線の種類は用いるパラメータの違いを表しており，prop $\w_\mathrm{Id}$，prop $\w_\mathrm{De}$，prop $\w_\mathrm{Un}$の場合は$p$，rank constの場合は$r$毎に異なる種類の線となっている．
In the case of Id, Obs, and Uni, the results corresponding to the different values of $p$ are shown with different line types.
Similarly, in the case of RC, the results corresponding to the different target ranks $r$ are shown with different line types.

%Figure \ref{fig:o3r4}はテンソルのサイズを$40\times 40\times 40$，ランクを$[4, 4, 4]$としたときの結果である．
Fig. \ref{fig:o3r4} is the result when the size of the original tensor is $40\times 40\times 40$ and the rank is $[4, 4, 4]$.
%Figure \ref{fig:o3r4}内のグラフ(a)-(d)から，以下のことが読み取れる．
From the graphs (a)-(d) in Fig. \ref{fig:o3r4}, one can see that
\begin{itemize}
    %\item prop $\w_\mathrm{Id}$では，$p$を変化させてもほとんど性能に変化はなく，$\alpha$の変化による性能の劣化は$p=1$が最も遅い．
    \item In the case of Id, if we choose $\alpha<2$, the choice of $p$ does not have much effect on the performance. 
    The slowest degradation of performance due to the change in $\alpha$ is obtained at $p=1$.
    %\item prop $\w_\mathrm{De}$の場合は，(a)-(d)のすべてで$p=1/2$が最も良い結果となった．この結果は先行研究での報告と一致している．（schatten系の文献引く）
    \item In the case of Obs, $p=1/2$ shows the best result across  (a)-(d). These results are consistent with the results in previous studies \cite{Kong2018,Xie2016,Zhang2019,Zha2018}. 
    %\item $p$の値によらず，すべての場合で$\w_\mathrm{Un}$を用いるよりも$\w_\mathrm{Id}$または$\w_\mathrm{De}$を用いた場合のほうが同じかより高い性能を示した．
    \item Regardless of $p$, the performance of Id and Obs is the same or better than that of Uni in all cases.
    %\item rank constは，すべての場合で$r$がテンソルランクと一致しているとき以外，非常に悪い結果となった．
    \item In all cases, RC shows the worst performance unless we can choose the correct rank $r$.
\end{itemize}

%また，Figure. \ref{fig:o3r4}と同様にテンソルのサイズを$40\times 40\times 40$とし，ランクを$[5, 5, 5]$に変えたときの結果をFigure. \ref{fig:o3r5}に示す．
Fig. \ref{fig:o3r5} shows the results when we only change the tensor rank to $[5,5,5]$. 
Note that the size of the tensor is still $40\times 40\times 40$.
%このときも，Figure. \ref{fig:o3r4}の場合と同様に上に示したような傾向が見られた．
The results show a similar trend as in the case of Fig. \ref{fig:o3r4}.
%Figure. \ref{fig:o3r4}，\ref{fig:o3r5}の結果から，weightingとSchatten-$p$の関係や，提案アルゴリズムとランク制約付き最小化の性能差に関しては，3階テンソルの場合ランクによらず同様の傾向が見られると考えられる．
From the results in Figs. \ref{fig:o3r4} and \ref{fig:o3r5}, for the third-order tensor, we can conclude that the effect of the choice of the weights, $p$, and the algorithms on performance is rank-independent.

%Figure. \ref{fig:o3r4}，\ref{fig:o3r5}で見られた傾向が，テンソルの階数を変化させたときにどのように変わるかを明らかにするために．4階テンソルを用いた実験を行った．
To reveal the impact of the changes in the order of the tensor on the common trend in Figs. \ref{fig:o3r4} and  \ref{fig:o3r5}, we performed experiments on the 4th-order tensor.
%結果をFigure. \ref{fig:o4r2}，\ref{fig:o4r3}に示す．
The results are shown in Figs. \ref{fig:o4r2} and \ref{fig:o4r3}.
%Figure. \ref{fig:o4r2}，\ref{fig:o4r3}は原テンソルのサイズを共に$16\times 16\times 16\times 16$，ランクをそれぞれ[2, 2, 2, 2]，[3, 3, 3, 3]とした．
In Figs. \ref{fig:o4r2} and \ref{fig:o4r3}, the sizes of the original tensors are both $16\times 16\times 16\times 16$, and the ranks are [2, 2, 2, 2] and [3, 3, 3, 3],  respectively.
%4階テンソルの場合も，ランクを変化させたとしても各グラフの傾向に変化はなかった．
In the case of the 4th-order tensor, there was no change in the common trend of each graph when the rank was varied.
%また，Figure. \ref{fig:o3r4}，\ref{fig:o3r5}と比較しても同様の傾向が見られた．
The same trend is observed in comparison with Figs. \ref{fig:o3r4} and \ref{fig:o3r5}.
%これらのことから，Figure. \ref{fig:o3r4}より得られた重み付けと$p$乗の関係や，提案アルゴリズムとランク制約付き最小化の性能差は，原テンソルの階数，ランクによらず一定の法則があるといえる．
From these observations, we can say that the relationship between the weighting and  Schatten-$p$ extension and the performance gap between the proposed algorithm and the rank-constrained minimization that we revealed is a law and is independent of  the tensor rank and order.

From these results, we can conclude that
\begin{itemize}
    %\item オラクルに近い高品質な重みを得られる場合$p=1$で十分である．
    \item It is sufficient to use $p=1$ if weights that are close to the ideal weights can be estimated in some way.
    %\item 劣化した特異値のみから重みを決定する場合は$p$を小さく設定したほうが良い．
    \item It is better to set a small $p$ value if the weights estimated from the degraded singular values are not reliable.
    %\item ランク制約を用いた単純な手法は，制約に用いるランクに非常に敏感であり，ランクを正しく推定できない限り，提案アルゴリズムのような複雑な手法を上回らない．
    \item Simple methods using rank constraints are very sensitive to the choice of ranks  used for the constraints and cannot outperform complex methods like the proposed algorithm unless one can correctly estimate the original ranks.
\end{itemize}
%最後だけintroまんま感あるからどうにかしたい
%subsec分けしたほうが良さそう要相談

%最後はintroで出たnatural questionの「weightedとpは相乗的か？片方がもう一方を包括するのか？」「単純なrank constが勝つことはないのか？」に答えるようにする．
%考察箇条書き
%oracle weightのときはp=1でいい←すべての結果でoracle weightはpを変化させても最高性能は変わらず，αを変化させても性能の劣化が始まるのが一番遅いため．また，非凸性が比較的マシで安定しやすいため．（安定しやすいはデータ出せないので書かなくてもいいかも）
%observation weightのときはp=1/2←ずべての結果でobserv weightはp＝1/2がp=2/3と比べて同程度または高性能なため．また，この結果は先行研究での報告と一致している．
%rankはrをちゃんと決めないと弱い←すべての結果でr=原テンソルランクの場合以外weight+schatten-pに性能が劣っているため．原テンソルランクを求めるのは非常に難しい．
%weightingとschatten-pの性能への寄与度
    %常にweightingはつねに入れたほうが強い．（pの値によらず）
    %observ weightの場合は殆どの場合weightingよりもschatten-pのほうが効果が強い．
    %強さはobserv weightのときはw+p＞p＞w＞なにもなし，oracle weightのときw+p＝w＞p＞なにもなし．
%生成過程ごととか観測過程ごとにも考察入れたほうがいいのかな？とりあえず今出てるの書いてから考える．

%メモ：たしかグラフの凡例のrが順番間違えてるので確認\&修正．

\section{Conclusion}
%やったことをまとめる
%introで出たnatural questionの「weightedとpは相乗的か？片方がもう一方を包括するのか？」「単純なrank constが勝つことはないのか？」に答えるようにする．
%↑の結論に到達するために色々積み上げていく．
%なにをやったか
%本論文では，特異値に対する重み付けとp乗の効果の間の関係性を明らかにするため，これらを組み合わせた一般的なテンソル復元モデルと，これを解くアルゴリズムを提案した．（inrtoママ）
In this paper, to reveal the relationships between the weighting and Schatten-$p$ extension, we propose a general tensor recovery model that combines them and propose an algorithm to solve it.

%実験でなにが得られたか
%提案アルゴリズムを用いた人工データに対する実験により，様々なシチュエーションにおいて，weight，pの有無が性能にどう影響するか明らかとなった
From the experiments with artificial data using the proposed algorithms, the effect of the recovery performance in the  presence or absence of the weighting and the Schatten-$p$ extension for various situations is determined.

%まとめの主張
%結果として，単純なrank constを用いた手法では，制約に用いるランク$r$を適切に決定できない限り，提案アルゴリズムのような複雑な手法を上回ることはなかった．
Consequently, the simple rank-constrained minimization method cannot outperform complex methods such as the proposed algorithm unless the rank $r$ used in the constraint is chosen properly.
%また，WTSPNにおけるweightingとpの関係はweightの決定方法によって異なり，oracleなweightの場合weightを導入すればpは必要ないが，観測から求めたweightの場合weightとpの効果が相乗的であることがわかった．
The relationships between the weighting and Schatten-$p$ extension in WTSPNs vary with the degree to which we can estimate the ideal weights.
The Schatten-$p$ extension does not affect the performance if the ideal weight is available.
On the other hand, the effect of the Schatten-$p$ and the weighting on singular values is synergistic if we need to determine the weights from heavily degraded observations.

\begin{figure}
    \centering
    \includegraphics[width=\linewidth]{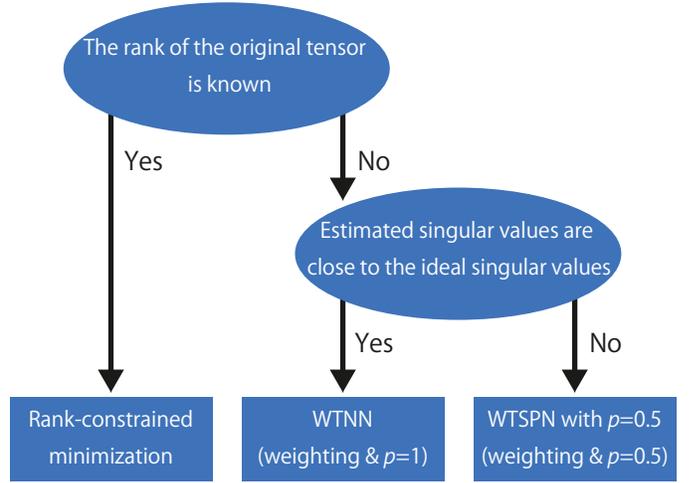}
    \caption{The flowchart for determining a method and parameter for the low-rank tensor recovery problem.\label{fig:fc}}
\end{figure}

%最終的にどう手法を決めるべきか
Our conclusion is summarized in the flowchart in Fig. \ref{fig:fc}, where ``weighting'' indicates that using the weights $\w$, which are determined based on the estimates of the singular values of the unfolded original tensor.
This flowchart implies that if we can access limited information about the rank (or the singular values) of the original tensor, we need to use complex methods to obtain good results.

\appendix
\section{Algorithm for rank-constrained minimization}
%sec2でEq.5のrank constrained minimizationはADMMで解けると言った．
In Section \ref{sec:pre}, we mentioned that we can solve Eq. (\ref{eq:rankinneq+L2fid}) efficiently by using ADMM, although we did not show a specific algorithm.
%this sectionではEq.(5)を解く具体的なアルゴリズムについて解説する．
The algorithm for solving Eq. (\ref{eq:rankinneq+L2fid}) is shown in Algorithm \ref{al:ADMM4rcm}.

%何行目がこれこれこうでproxはこうとか
The objective function in line 5 of Algorithm\ref{al:ADMM4rcm} is
\begin{equation}
\label{eq:pro2rc}
    \proj_{\{\X|\mathrm{rank}(\X)\leq\hat{r}_m\}}(\unfold_m(\cX^{(k+1)})+\Z_{1,m}^{(k)}), 
\end{equation}
%このアルゴリズムでsec4の実験もやったよ
which is a nonconvex function because the set  $\{\X|\mathrm{rank}(\X)\leq\hat{\r}_m\}$ is a nonconvex set.
However, one of the solutions of Eq. (\ref{eq:pro2rc}) can be obtained as
\begin{equation}
    \label{eq:proxofrc}
    \U\mathrm{T}_{\hat{r}_m}(\Sigma)\V^{\top},
\end{equation}
where $\U\Sigma\V^\top=\unfold_m(\cX^{(k+1)})+\Z_{1,m}^{(k)}$ is an SVD and $\mathrm{T}_{\hat{r}_m}(\cdot)$ is a truncation operator.
The truncation operator for a rectangular diagonal matrix $\Y$ $\mathrm{T}_{r}(\Y)$ is defined as
\begin{equation}
    \label{eq:truncate}
    \mathrm{T}_{r}(\Y)_{i,i} = \left\{ \begin{array}{ll}
    \Y_{i,i} & (i\geq r) \\
    0 & (\textrm{otherwise})
  \end{array}\right.
  .
\end{equation}

\begin{algorithm}[t]
\caption{Algorithm for rank-constrained minimization}
\label{al:ADMM4rcm}
{%\small
\begin{algorithmic}[1]
\REQUIRE $\cY$, $\lambda$, $r_1,\cdots,r_N$
\STATE{\textrm{Initialize} $\Y_{1,m}^{(0)}=\unfold_m(\cY)$, $\cY^{(0)}_2=\cY$, $\Z_{1,m}^{(0)}=0$, $\Y_{1,m}^{(0)}$, $\cZ_2^{(0)}=0$}
\WHILE{\textit{A stopping criterion is not satisfied}}
	\STATE $\cX^{(k+1)}=\argmin_{\cX}\frac{1}{2}\sum_{m=1}^N\|\Y_{1,m}^{(k)}-\unfold_m(\cX)-\Z_{1,m}^{(k)}\|^2_2$\\
	$+\lambda\|\cY_2^{(k)}-\mathrm{A}_\Omega(\cX)-\cZ_2^{(k)}\|^2_2$
	\FOR{$m = 1$ to $N$}
	    \STATE $\Y_{1,m}^{(k+1)}=\proj_{\{\X|\mathrm{rank}(\X)\leq\hat{r}_m\}}(\unfold_m(\cX^{(k+1)})+\Z_{1,m}^{(k)})$
	    \STATE $\Z_{1,m}^{(k+1)}=\Z_{1,m}^{(k)}+\unfold_m(\cX^{(k+1)})-\Y_{1,m}^{(k+1)}$
	\ENDFOR
	\STATE $\cY_2^{(k+1)}=\argmin_{\cX}\lambda\|\A_\Omega(\cX)-\cY\|^2_2+\frac{1}{2}\|\cX^{(k+1)}+\cZ_2^{(k)}-\cX\|_F^2$
	\STATE $\cZ_2^{(k+1)}=\cZ_2^{(k)}+\mathrm{A}_{\Omega}(\cX^{(k+1)})-\cY_2^{(k+1)}$
	\STATE $\lambda = 0.99\lambda$
	\STATE $k=k+1$
\ENDWHILE
\ENSURE $\cX^{(k)}$
\end{algorithmic}
}
\end{algorithm}

\section*{Acknowledgment}

The authors would like to thank...

\ifCLASSOPTIONcaptionsoff
  \newpage
\fi

%\bibliography{refs}
%\bibliographystyle{ieeetr}
\def\ringaccent#1{{\accent23 #1}}



%\newpage


%\vfill


\end{document}